\documentclass[a4paper,11pt]{article}
\usepackage{heppub}
\pdfoutput=1
\usepackage{graphicx}
\usepackage{subcaption}
\usepackage{lipsum}
\usepackage{booktabs}
\usepackage{tabularx}
\usepackage{tikz}
\usepackage{pgfplots}
\usetikzlibrary{positioning,shapes,trees,arrows,graphs,decorations,calc}
\usepackage{url}

\usepackage[inline]{enumitem}

\usepackage{amssymb}
\usepackage{amsmath}
\usepackage{hyperref}
\usepackage{xcolor}
\usepackage{placeins} % in order to use floatbarrier
\usepackage[normalem]{ulem}
\usepackage{xspace}

\newcommand{\ord}[1]{\mathcal{O}(#1)}

\newcommand{\df}{\mathrm{d}}

\newcommand{\as}{\alpha_{s}}

\newcommand{\Tau}{\mathcal{T}}

\newcommand{\Obs}{\mathcal{O}}

\newcommand{\GeV}{\,\mathrm{GeV}}

\newcommand{\nn}{\nonumber}

\newcommand{\cut}{\mathrm{cut}}

\newcommand{\FO}{\mathrm{FO}}

\newcommand{\NNLO}{\mathrm{NNLO}}

\newcommand{\geneva}{\textsc{Geneva}\xspace}

\newcommand{\nnlojet}{\textsc{nnlojet}\xspace}

\newcommand{\scetlib}{\textsc{SCETlib}\xspace}

\newcommand{\openloopsTwo}{\textsc{OpenLoops2}\xspace}

\def\thetaPSiso{\ensuremath{\Theta^{\mathrm{PS}}}\xspace}

\def\thetaProj{\ensuremath{\Theta^{\mathrm{PS}}_{\mathrm{proj}}\xspace}}

\def\thetaBarProj{\ensuremath{\overline{\Theta}^{\mathrm{PS}}_{\mathrm{proj}}\xspace}}

\newcounter{bla}

\newcommand{\rescaletwoplots}{0.49\textwidth}

\makeatletter
\g@addto@macro\bfseries{\boldmath}
\makeatother

\DeclareMathAlphabet\mathbfcal{OMS}{cmsy}{b}{n}

\title{NNLO predictions with nonlocal subtractions and fiducial power corrections in GENEVA}

\author[a]{Simone Alioli,}
\author[a,b]{Georgios Billis,}
\author[c]{Alessandro Broggio}
\author[a]{and Giovanni Stagnitto}
\affiliation[a]{Universit\`a degli Studi di Milano-Bicocca \& INFN
  Sezione di Milano-Bicocca, \\ Piazza della Scienza 3, Milano 20126,
  Italy}
\affiliation[b]{Paul Scherrer Institut, CH-5232 Villigen PSI, Switzerland}
\affiliation[c]{Faculty of Physics, University of Vienna, Boltzmanngasse 5, A-1090 Wien, Austria}

\abstract{We present the implementation of next-to-next-to-leading
  order (NNLO) QCD fully-differential corrections within the \geneva
  framework, for both colour-singlet and colour-singlet+jet processes at hadron colliders, by
  employing a nonlocal subtraction approach. In particular, we discuss the
  implementation details and the challenges that arise when utilizing
  a dynamical infrared cutoff parameter.
  Additionally, we combine the subtraction with the projection-to-Born method
  in order to include fiducial power corrections. As a test case,
  we provide predictions for Drell-Yan and $Z$+jet production at the LHC,
  using $N$-jettiness as resolution variable.
  We validate the NNLO corrections of \geneva against \nnlojet finding
  excellent agreement. Finally, we discuss how to extend our method
  to calculate the N$^3$LO QCD fully-differential corrections to colour-singlet
  production at hadron colliders.}

\begin{document}

\begin{flushright}
	PSI-PR-25-08\\
	UWThPh 2025-14
\end{flushright}

%\date{\today}

\maketitle

\section{Introduction}

The High Luminosity Large Hadron Collider (HL-LHC) is set to begin operations by the end of the decade, and will provide the LHC experiments with significantly larger data samples, up to an order of magnitude greater than the current baseline program. This will present a unique opportunity to achieve unprecedentedly precise measurements across the entire phase space, while also expanding the potential to detect deviations from Standard Model (SM) predictions across a wide range of processes and observables. In particular, the increased statistics will allow for highly accurate measurements of the high-energy tails of distributions, which are especially sensitive to new physics effects.

In order to ensure that the theoretical predictions match the precision of the experimental measurements at the LHC, it is essential to achieve at least next-to-next-to-leading order (NNLO) accuracy in the perturbative expansion in the strong coupling $\alpha_s$ when evaluating partonic cross sections of crucial benchmark processes, such as the production of electroweak bosons ($W^\pm,Z$)~\cite{CMS:2024myi,ATLAS:2025ede,ATLAS:2024nrd,ATLAS:2025hhn,CMS:2025sgv}.

Production processes of SM bosons in association with a hadronic jet are among
the most frequently occurring reactions at hadron colliders and retain a large
event rate. At they same time, they also offer the possibility to study
more precisely the transverse momentum ($q_T$) spectrum of SM bosons, which is primarily
caused by the recoil against emitted partons and depends on QCD dynamics.
In particular, the production of a $Z$ boson in association with a jet is a key process in probing SM physics, for
example in fitting Parton Distribution Functions (PDFs) and measuring the
strong coupling constant $\alpha_s$~\cite{Camarda:2022qdg, ATLAS:2023lhg}. Furthermore, when the $Z$ boson decays into invisible neutrinos, it is an important background for searches of physics
beyond the SM that involve missing energy.
Similarly, the Higgs+jet process allows for searches of non-standard Higgs couplings while
the $W$+jet process serves as a background for dark matter searches~\cite{CMS:2021snz, ATLAS:2021kxv}. Precise predictions of the $W$-boson’s transverse momentum are also crucial for accurate
$W$-boson mass determinations~\cite{ATLAS:2017rzl, LHCb:2021bjt, CDF:2022hxs, CMS:2024lrd, ATLAS:2024erm}.
In addition, NNLO predictions for colour-singlet+jet processes are fundamental ingredients for fully-differential calculations of $Z$/Higgs/$W$ production at next-to-next-to-next-to-leading order (N$^3$LO) accuracy~\cite{Caola:2022ayt,Chen:2022lwc,Campbell:2023lcy,Chen:2021vtu,Chen:2022cgv,Neumann:2022lft,Cieri:2018oms,Chen:2021isd,Billis:2021ecs,Billis:2024dqq}.

The evaluation of fully differential NNLO corrections requires the implementation of a subtraction or a slicing method to remove the infrared (IR) divergences that appear at intermediate stages of the calculation.
In the last decade, NNLO predictions for processes with final-state jets at the LHC were obtained by different groups using local subtraction methods~\cite{Boughezal:2013uia} such as the antenna subtraction scheme~\cite{Gehrmann-DeRidder:2005btv}, the \textsc{Stripper} scheme~\cite{Czakon:2010td,Czakon:2011ve} as well as $N$-jettiness ($\Tau_N$) slicing~\cite{Boughezal:2015ded,Neumann:2022lft}.

Nonlocal subtraction methods at NNLO which utilize as slicing variable either the transverse momentum ($q_T$) of the colour-singlet system \cite{Catani:2007vq, Grazzini:2017mhc, Campbell:2022gdq}, or $N$-jettiness \cite{Gaunt:2015pea, Boughezal:2015dva, Boughezal:2015ded, Boughezal:2016wmq, Campbell:2019dru}, or a process-specific slicing parameter~\cite{Gao:2012ja, Liu:2018gxa},
can be further refined in the presence of fiducial cuts by the combination with the Projection-to-Born (P2B) method \cite{Cacciari:2015jma}, as shown for the first time in \refscite{Ebert:2019zkb} and implemented up to NNLO in \cite{Campbell:2024hjq}. This improvement allows to include fiducial power corrections (FPCs) below the IR cutoff parameter $\Tau_{\delta}$ for the observables that are sensitive to that. The inclusion of these effects allows to obtain a better agreement with local subtraction methods.

In this paper we present a novel implementation of nonlocal subtractions in combination
with the P2B method for colour-singlet and colour-singlet+jet processes at the LHC.
At variance with previous implementations which rely on a pure slicing approach,
here we implemented a genuine nonlocal subtraction term which is integrated together with the
NLO differential cross section for a resolved emission above $\Tau_\delta$.
To the best of our knowledge, this is the first implementation of a genuinely nonlocal
subtraction. This requires the introduction of normalized splitting functions that renders
the subtraction terms differential in the higher multiplicity phase space of the
resolved emission while preserving its dependence on the resolution variable.
The primary challenge of this approach, when using a dynamic
$\Tau_{\delta}$ parameter, lies in the need for a mapping of the splitting
functions that preserves the variables which define the dynamical cutoff.

While the formulas that we derive are independent of specific resolution variables and can be
applied to any suitable variable that correctly captures the IR regions of QCD in the limit where
this variable approaches zero, in this work we present, as an example, results for the $Z$
and $Z$+jet production processes by employing $N$-jettiness subtractions \cite{Gaunt:2015pea}
in combination with the P2B method.
Our implementation of nonlocal $N$-jettiness subtractions relies on the fixed
order expansion up to relative  $\mathcal{O}(\alpha_s^2)$ of leading power
N$^3$LL resummation formulas for zero-jettiness  ($\Tau_0$)
\cite{Alioli:2023har,Billis:2019vxg} and 1-jettiness  ($\Tau_1$)
\cite{Alioli:2023rxx}, both defined in the frame where the colour-singlet system has zero rapidity.
In the $0$-jettiness case, we also supplement the leading power approximation
with the leading next-to-leading power (NLP) logarithms which were computed in
\cite{Moult:2016fqy} and provide an improvement in the numerical efficiency
and stability of subtractions, which allows for a larger
$\Tau_{\delta}$ parameter. Unfortunately the
corresponding leading-logarithm NLP contributions are not yet known for $1$-jettiness.
We also point out that for the $Z$+jet calculation we employ the $1$-jettiness definition derived in \cite{Alioli:2023rxx}, which is based on an exclusive clustering procedure. This means that we recursively cluster together emissions using the  $\Tau_1$ metric until we are left with one jet (see discussion in \cite{Alioli:2023rxx}) and we don't rely on external jet clustering algorithms. We stress that this definition is different from the one used in \cite{Boughezal:2015dva,Boughezal:2015ded,Boughezal:2015aha,Campbell:2019gmd}, where the jet axis is determined a priori by employing an inclusive jet clustering procedure.

The results of this paper, besides providing independent NNLO calculations,
are also crucial for the development of the \textsc{Geneva} Monte Carlo (MC)
event generator
\cite{Alioli:2015toa,Alioli:2019qzz,Alioli:2020fzf,Alioli:2020qrd,Alioli:2021egp,Cridge:2021hfr,Alioli:2022dkj,Alioli:2023har}. Indeed
they eliminate the potential need for a reweighting procedure, typically
carried out with an external NNLO-accurate program, by directly providing
instead the NNLO predictions for the $N$-jet cross section.

The manuscript is organized as follows. In \sec{theory} we introduce the formulas which combine a nonlocal subtraction scheme with the P2B method in order to capture the fiducial power corrections. This formalism is general and can be applied to any suitable jet resolution variable and processes with any number of final-state jets.
In \sec{genimpl} we derive the relevant cross section formulas for generic $N$-jettiness variables in a form that is suitable for the implementation in the \geneva framework.
In \sec{implementation} we specify the subtraction formulas for colour-singlet and colour-singlet+jet processes and provide numerical results for $Z$ and $Z$+jet using
$N$-jettiness subtractions in combination with the P2B method, comparing our results to \nnlojet~\cite{NNLOJET:2025rno,Berends:1988yn,Bern:1997sc,Boughezal:2010mc,Buckley:2014ana,Currie:2013vh,Daleo:2006xa,Daleo:2009yj,Garland:2001tf,Garland:2002ak,Gauld:2017tww,Gauld:2021pkr,Gehrmann:2002zr,Gehrmann:2011ab,Gehrmann:2011wi,Gehrmann-DeRidder:2004ttg,Gehrmann-DeRidder:2005alt,Gehrmann-DeRidder:2005btv,Gehrmann-DeRidder:2005svg,Gehrmann-DeRidder:2007foh,Gehrmann-DeRidder:2012too,Gehrmann-DeRidder:2015wbt,Gehrmann-DeRidder:2016cdi,Gehrmann-DeRidder:2016jns,Gehrmann-DeRidder:2019avi,Gehrmann-DeRidder:2023urf,Hagiwara:1988pp}.
In \sec{n3lo} we discuss the extension of our approach to obtain N$^3$LO fully-differential predictions.
We finally draw our conclusions in \sec{conclusions}.

\section{Theoretical framework}
\label{sec:theory}

We calculate the NNLO QCD corrections ($\delta$NNLO) to a generic
observable $\Obs$, where the subtraction terms and their integrated
counterparts are understood and not written out explicitly.
For ease of notation we also omit flavour indices in the cross sections, which are implicitly  given relative to a partonic channel.
Summing over all relevant partonic channels we get
the total NNLO corrections $\delta$NNLO, and adding it to the corresponding NLO results
we obtain the total NNLO cross sections and distributions.
Since the calculation of NLO corrections for a generic observable are easily obtained with fully-local subtraction methods, we will not discuss them further in this paper.

Our starting formula for the NNLO corrections is
\begin{align}
\Obs_{\delta\textrm{NNLO}} (\Phi_N)  = & W(\Phi_N) \Obs(\Phi_N) + \int \frac{\df \Phi_{N+1}}{\df \Phi_{N}} R\!V(\Phi_{N+1}) \Obs(\Phi_{N+1}) \\ &+ \int \frac{\df \Phi_{N+2}}{\df \Phi_{N}} R\!R(\Phi_{N+2}) \Obs(\Phi_{N+2})\nn\\
\equiv & \frac {\df \sigma_N^{\delta\textrm{NNLO}}}{\df \Phi_N} \Obs(\Phi_{N+X})\nn\,,
\end{align}
where in  the last line we have defined the NNLO correction
cross section  at fixed underlying-Born\footnote{Throughout this paper, we adopt the point of view of dressing a Born configuration with higher-order splittings, rather than starting from real emission diagrams and map them to a Born kinematics. The notation $\df \Phi_{N+k}/\df \Phi_{N}$ is thus a shorthand notation for the radiation phase space with $k$ additional emissions at fixed underlying-Born kinematics. For example
\begin{align}
  \int \frac {\df \Phi_{N+1}}{\df \Phi_{N}}   = & \int \df \Phi_{N+1}\ \delta \left( \widetilde \Phi_N(\Phi_{N+1}) - \Phi_N  \right)\,, \nn
\end{align}
and $\widetilde \Phi_N(\Phi_{N+1})$ represent any possible projection of the
$N+1$-body phase-space on the underlying-Born one $\Phi_N$.}
kinematics $\df \sigma_N^{\delta\textrm{NNLO}} / \df \Phi_N$ and the notation
$\Obs(\Phi_{N+X})$ indicates that the observable is evaluated at the exact
kinematic configuration for each one of the NNLO contributions: double virtual
$W(\Phi_N)$, real-virtual $R\!V (\Phi_{N+1})$ and double real $R\!R(\Phi_{N+2})$.

The formula above implies that for processes which are already divergent at the Born level, a set of process-defining cut must be applied to all contributions.
A typical example in the process $pp \to Z$+jet would be a cut on the transverse momentum ($q_T$) of the $Z$-boson
or on  zero-jettiness ($\Tau_0$).

One can now add and subtract the observable evaluated on the projected Born  phase space $\Obs(\Phi_N)$ and obtain the equivalent formula
\begin{align}
\label{eq:p2b_prototypical_form}
\Obs_{\delta\textrm{NNLO}} (\Phi_N)   = &
 \frac {\df \sigma_N^{\delta\textrm{NNLO}}}{\df \Phi_N}  \Obs(\Phi_{N}) + \frac {\df \sigma_N^{\delta\textrm{NNLO}}}{\df \Phi_N} \bigg[ \Obs(\Phi_{N+X}) - \Obs(\Phi_{N})\bigg]  \, .
\end{align}
At this point, we observe that the double virtual term  never contributes to the second addend of \eq{p2b_prototypical_form},
since its contribution is multiplied by the difference between the observable evaluated on the same $\Phi_N$ configuration, which is zero.
Therefore, without doing any approximation, the second addend is exactly replaced by the cross section of the NLO corrections for the process with one extra
parton, integrated over the additional radiation phase space at fixed the underlying-Born kinematics, i.e.
\begin{align}
  \label{eq:eq4}
\Obs_{\delta\textrm{NNLO}} (\Phi_N)  = &
  \frac {\df \sigma_N^{\delta\textrm{NNLO}}}{\df \Phi_N}  \Obs(\Phi_{N}) +\int \frac {\df \Phi_{N+1}}{\df \Phi_{N}} \frac{ \df \sigma_{N+1}^{\delta\textrm{NLO}}}{\df \Phi_{N+1}} \bigg[ \Obs(\Phi_{N+X}) - \Obs(\Phi_{N})\bigg] \,.
\end{align}
\Eq{eq4} is the essence of the projection-to-Born (P2B) method. In
order to be practically applied, however, it still requires
the knowledge of the exact NNLO corrections evaluated on the underlying-Born
kinematics (first addend of \eq{eq4}).
In the following we will show how this requirement can be circumvented by
leveraging on a subtraction approach, which relies on an approximation of this term, that
 can be systematically improved until the desired accuracy is reached.
In order to proceed we choose a generic resolution variable $\Tau_N$ and split the first addend of \eq{eq4} by means of a $\Tau_N^{\textrm{cut}}$.
Any resolution variable can be used, provided that  $\Tau_N (\Phi_N) \equiv 0$ and that $\Tau_N (\Phi_{N+M}) \to 0 $ in any infrared (IR) limit when the $\Phi_{N+M}$ configuration approaches  $\Phi_N$.
Typical examples of such resolution variables are the transverse momentum of the colour singlet $q_T$ for colour-singlet production or $N$-jettiness for processes that also include jets.
Using the resolution cut, we can therefore rewrite
\begin{align}
\label{eq:eq5}
\Obs_{\delta\textrm{NNLO}} (\Phi_N)  = &  \frac {\df \Sigma_N^{\delta\textrm{NNLO}}}{\df \Phi_N} \big( \Tau_N^{\textrm{cut}}\big)   \,\Obs(\Phi_{N})\   \\
& + \int \frac {\df \Phi_{N+1}}{\df \Phi_{N}} \frac{ \df \sigma_{N+1}^{\delta\textrm{NLO}}}{\df \Phi_{N+1}} \  \Obs(\Phi_{N})  \ \theta\big(\Tau_N(\Phi_{N+X}) > \Tau_N^{\textrm{cut}} \big) \nn \\
&  + \int \frac {\df \Phi_{N+1}}{\df \Phi_{N}} \frac{ \df \sigma_{N+1}^{\delta\textrm{NLO}}}{\df \Phi_{N+1}} \bigg[ \Obs(\Phi_{N+X}) - \Obs(\Phi_{N})\bigg]\, ,  \nn
\end{align}
where we have defined the cumulant below $\Tau_N^{\textrm{cut}}$ as
\begin{align}
\frac {\df \Sigma_N^{\delta\textrm{NNLO}}}{\df \Phi_N} \big( \Tau_N^{\textrm{cut}}\big) & = W(\Phi_N)  + \int \frac{\df \Phi_{N+1}}{\df \Phi_{N}} R\!V(\Phi_{N+1}) \theta\big(\Tau_N(\Phi_{N+1}) < \Tau_N^{\textrm{cut}} \big)\nn \\
&\quad + \int \frac{\df \Phi_{N+2}}{\df \Phi_{N}} R\!R(\Phi_{N+2}) \theta\big(\Tau_N(\Phi_{N+2}) < \Tau_N^{\textrm{cut}} \big) \, \nn \\
& = \int_0^{\Tau_N^\cut}\!\!\!\! \df \Tau_N \ \frac {\df \sigma_{N}^{\delta\textrm{NNLO}}}{\df \Phi_N\, \df \Tau_N}\,,
\end{align}
and the $\Tau_N$ spectrum  as
\begin{align}
\frac {\df \sigma_{N}^{\delta\textrm{NNLO}}}{\df \Phi_N\, \df \Tau_N} & = W(\Phi_N) \delta(\Tau_N) + \int \frac{\df \Phi_{N+1}}{\df \Phi_{N}} R\!V(\Phi_{N+1}) \delta\big(\Tau_N(\Phi_{N+1}) - \Tau_N \big)\nn \\
&\quad + \int \frac{\df \Phi_{N+2}}{\df \Phi_{N}} R\!R(\Phi_{N+2}) \delta\big(\Tau_N(\Phi_{N+2}) - \Tau_N \big) \,.
\end{align}
In the second line of \eq{eq5} we have again replaced the NNLO correction cross section above $\Tau_N^{\textrm{cut}}$ by the cross section of the NLO corrections for one extra parton, integrated over the additional phase space, since the two are identical by definition.
We can now add and subtract a ``subtraction'' cross section
\begin{align}
  \frac {\df \Sigma_{N,\textrm{sub.}}^{\delta\textrm{NNLO}}}{\df \Phi_N} \big(
  \Tau_N^{\textrm{cut}}\big ) & = \int_0^{\Tau_N^\cut}\!\!\!\! \df \Tau_N \ \frac {\df \sigma_{N,\textrm{sub.}}^{\delta\textrm{NNLO}}}{\df \Phi_N\, \df \Tau_N}\,,
\end{align}
and  obtain
\begingroup
\allowdisplaybreaks
  \begin{align}
    \label{eq:tauNcut}
\Obs_{\delta\textrm{NNLO}} (\Phi_N)  = &  \frac {\df \Sigma_{N,\textrm{sub.}}^{\delta\textrm{NNLO}}}{\df \Phi_N} \big( \Tau_N^{\textrm{cut}}\big)   \,\Obs(\Phi_{N})  \\
&  + \int \df \Tau_N \left[ \frac {\df \sigma_{N}^{\delta\textrm{NNLO}}}{ \df \Phi_N\, \df \Tau_N} - \frac {\df \sigma_{N,\textrm{sub.}}^{\delta\textrm{NNLO}}}{\df \Phi_N\, \df \Tau_N}\right] \  \Obs(\Phi_{N})\ \theta\big(\Tau_N < \Tau_N^{\textrm{cut}}\big) \nn \\
&  + \int \frac {\df \Phi_{N+1}}{\df \Phi_{N}} \frac{ \df \sigma_{N+1}^{\delta\textrm{NLO}}}{\df \Phi_{N+1}} \  \Obs(\Phi_{N})  \ \theta\big(\Tau_N(\Phi_{N+X}) > \Tau_N^{\textrm{cut}} \big) \nn \\
& + \int \frac {\df \Phi_{N+1}}{\df \Phi_{N}} \frac{ \df \sigma_{N+1}^{\delta\textrm{NLO}}}{\df \Phi_{N+1}} \bigg[ \Obs(\Phi_{N+X}) - \Obs(\Phi_{N})\bigg]\, .\nn
  \end{align}
  \endgroup
The ``subtraction'' cross section can be chosen to be any valid approximation of the exact
NNLO result. The minimum requirement is that it must include the leading-power (LP) singular contributions in $\Tau_N$, while it may also incorporate subleading-power corrections.
The crucial point is that it has to reproduce at least the same singular structure in the variable $\Tau_N$ of the full $\delta$NNLO.
We stress that the value of $\Tau_N^\cut$ is completely arbitrary up to
this point and  \eq{tauNcut} does not contain any approximation.
It is however useful to introduce a tiny IR cutoff $\Tau_\delta \ll \Tau_N^{\mathrm{cut}}$ which we use to split the second line of \eq{tauNcut} into
\begin{align}
  \label{eq:taudelta}
\Obs_{\delta\textrm{NNLO}} (\Phi_N)  = &   \frac {\df \Sigma_{N,\textrm{sub.}}^{\delta\textrm{NNLO}}}{\df \Phi_N} \big( \Tau_N^{\textrm{cut}}\big)   \,\Obs(\Phi_{N})  \\
&  + \int \df \Tau_N \left[ \frac {\df \sigma_{N}^{\delta\textrm{NNLO}}}{\df \Phi_N\, \df \Tau_N} - \frac {\df \sigma_{N,\textrm{sub.}}^{\delta\textrm{NNLO}}}{\df \Phi_N\, \df \Tau_N}\right] \  \Obs(\Phi_{N})\ \theta\big(\Tau_N < \Tau_\delta \big) \nn \\
&  + \int \df \Tau_N \left[ \frac {\df \sigma_{N}^{\delta\textrm{NNLO}}}{\df \Phi_N\, \df \Tau_N} - \frac {\df \sigma_{N,\textrm{sub.}}^{\delta\textrm{NNLO}}}{\df \Phi_N\, \df \Tau_N}\right] \  \Obs(\Phi_{N})\ \theta\big( \Tau_\delta <  \Tau_N < \Tau_N^{\textrm{cut}}\big) \nn \\
&  + \int \frac {\df \Phi_{N+1}}{\df \Phi_{N}} \frac{ \df \sigma_{N+1}^{\delta\textrm{NLO}}}{\df \Phi_{N+1}} \  \Obs(\Phi_{N})  \ \theta\big(\Tau_N(\Phi_{N+X}) > \Tau_N^{\textrm{cut}} \big) \nn \\
& + \int \frac {\df \Phi_{N+1}}{\df \Phi_{N}} \frac{ \df \sigma_{N+1}^{\delta\textrm{NLO}}}{\df \Phi_{N+1}} \bigg[ \Obs(\Phi_{N+X}) - \Obs(\Phi_{N})\bigg]\, .\nn
\end{align}
We now observe that due to the presence of the lower IR cutoff $\Tau_\delta$
that regulates any possible divergence in $\Tau_N$, the third line can again
be rewritten using the NLO corrections for one extra parton integrated
over the additional phase space.
The same can also be done  for the ``subtraction'' cross section, but here
one needs to introduce some splitting function
$P(\Phi_{N+1}) \propto \df \Phi_N \df \Tau_N / \df \Phi_{N+1}$ to make the ``subtraction'' cross section differential in the
$\Phi_{N+1}$ phase space. For a generic function $f(\Phi_N, \Tau_N)$  the splitting functions have to obey the normalization relation
\begin{align}
f(\Phi_N, \Tau_N) &=  \int  \frac {\df \Phi_{N+1}}{\df \Phi_{N}}  f(\Phi_N, \Tau_N)  P(\Phi_{N+1})\,,
\end{align}
that ensures that the subtraction remains local in the $\Tau_N$ variable.
Eventually we obtain
\begingroup
\allowdisplaybreaks
\begin{align}
\label{eq:first_option}
\Obs_{\delta\textrm{NNLO}} (\Phi_N)  = &  \frac {\df \Sigma_{N,\textrm{sub.}}^{\delta\textrm{NNLO}}}{\df \Phi_N} \big( \Tau_N^{\textrm{cut}}\big)   \,\Obs(\Phi_{N})  \\
&  + \int \df \Tau_N \left[ \frac {\df \sigma_{N}^{\delta\textrm{NNLO}}}{\df \Phi_N\, \df \Tau_N} - \frac {\df \sigma_{N,\textrm{sub.}}^{\delta\textrm{NNLO}}}{\df \Phi_N\, \df \Tau_N}\right] \  \Obs(\Phi_{N})\ \theta\big(\Tau_N  < \Tau_\delta \big) \nn \\
&  + \int  \frac {\df \Phi_{N+1}}{\df \Phi_{N}} \bigg [ { \frac{ \df \sigma_{N+1}^{\delta\textrm{NLO}}}{\df \Phi_{N+1}}}  - \frac {\df \sigma_{N,\textrm{sub.}}^{\delta\textrm{NNLO}}}{\df \Phi_N\, \df \Tau_N} \ P(\Phi_{N+1}) \, \theta\big( \Tau_N (\Phi_{N+1}) < \Tau_N^{\textrm{cut}}\big)  \bigg]  \nn \\
& \qquad \qquad \qquad \times  \Obs(\Phi_{N})  \ \theta\big(\Tau_N(\Phi_{N+X}) > \Tau_\delta \big)\nn \\
& + \int \frac {\df \Phi_{N+1}}{\df \Phi_{N}} \frac{ \df \sigma_{N+1}^{\delta\textrm{NLO}}}{\df \Phi_{N+1}} \bigg[ \Obs(\Phi_{N+X}) - \Obs(\Phi_{N})\bigg]\, .\nn
\end{align}
\endgroup
We remark that no approximation went in the derivation up to now, and all the
terms in the equation above are exactly calculable down to the value of
$\Tau_N$ reaching zero, because either the divergences are subtracted by an
approximate $\delta$NNLO correction as in the second line or by the difference of
the observables as in the last line (which ensures finiteness for any IR safe
observable by the KLN theorem).
By inspecting the third line of \eq{first_option} the complementary role of
the cuts on $\Tau_N$ becomes clear: $\Tau_N^\cut$ is an upper limit on the
range where the ``subtraction'' cross section is applied, while $\Tau_\delta$
acts as the true lower IR cutoff. When they are taken equal the
subtraction region shrinks to zero and one is back to the traditional slicing
approach.

The inclusion of the subtraction in the $\Tau_N$ spectrum is particularly delicate in
the presence of generating cuts, that serve to keep the cross section
finite over singular regions in the underlying-Born phase space. Because of these extra conditions, the
choice of the phase-space mappings used in the calculation of the NLO cross
section $\df \sigma_{N+1}^{\delta\textrm{NLO}} / \df \Phi_{N+1}$ or in the splitting functions $P(\Phi_{N+1})$ require particular attention.
Complete freedom over the choice of the mapping in NLO cross section is still guaranteed, provided that the mismatch generated by mappings that do not preserve the generating cuts is accounted for. Instead, only by preserving the generating cuts in the mappings used for the splitting functions one can ensure that the entirety of the allowed  phase space is correctly covered, and the correct predictions are obtained.
This will be further discussed in \sec{genimpl}.

At this point we are ready to discuss which terms to drop from \eq{first_option} and the corresponding approximation.
By dropping the second line we are neglecting the subleading  inclusive power corrections below the $\Tau_\delta$ tiny cutoff, and only those.
Indeed, since the observable is always evaluated on the underlying-Born kinematics in that line, there are no missing fiducial power corrections FPCs.
The FPCs below the $\Tau_\delta$ tiny cutoff are instead included by evaluating the NLO corrections for one extra jet down $\Tau_N = 0$,
as done in the last line of \eq{first_option}. This is possible because the divergences of the cross-section  are regulated by the difference in the observable.
Highlighting again that the only approximation is neglecting  the
inclusive power corrections below the infrared cutoff $\Tau_\delta$, our final
formula is
\begin{align}
  \label{eq:finalNNLO}
	\Obs_{\delta\textrm{NNLO}} (\Phi_N)  = &  \frac {\df \Sigma_{N,\textrm{sub.}}^{\delta\textrm{NNLO}}}{\df \Phi_N} \big( \Tau_N^{\textrm{cut}}\big)   \,\Obs(\Phi_{N})  \\
	&  + \int  \frac {\df \Phi_{N+1}}{\df \Phi_{N}} \bigg [ { \frac{ \df \sigma_{N+1}^{\delta\textrm{NLO}}}{\df \Phi_{N+1}}}  - \frac {\df \sigma_{N,\textrm{sub.}}^{\delta\textrm{NNLO}}}{\df \Phi_N\,\df \Tau_N} \ P(\Phi_{N+1}) \, \theta\big( \Tau_N (\Phi_{N+1}) < \Tau_N^{\textrm{cut}}\big)  \bigg]  \nn \\
	& \qquad \qquad \qquad \times  \Obs(\Phi_{N})  \ \theta\big(\Tau_N(\Phi_{N+X}) > \Tau_\delta \big)\nn \\
	& + \int \frac {\df \Phi_{N+1}}{\df \Phi_{N}} \frac{ \df
          \sigma_{N+1}^{\delta\textrm{NLO}}}{\df \Phi_{N+1}} \bigg[
          \Obs(\Phi_{N+X}) - \Obs(\Phi_{N})\bigg]\nn\\
      & + \textrm{neglected inclusive power corrections in }\Tau_\delta \nn\,.
\end{align}
The size of the neglected inclusive power corrections can be
controlled by the choice of the subtraction cross section used. The
simplest choice is to use the singular cumulant at relative order
$\alpha^2_{\textrm{s}}$ for
$ \df \Sigma_{N,\textrm{sub.}}^{\delta\textrm{NNLO}} / \df \Phi_N$,
i.e. including only the leading-power contributions. This can however
be ameliorated by adding further subleading power terms. For example,
including the subleading-power leading logarithms one can make the
inclusive power corrections in $\Tau_\delta$ in the last line smaller.
The choice of the subtraction cross section cumulant in the first line
determines the choice the subtraction spectrum appearing in the second
line. Ultimately this determines the associated nonsingular spectrum,
i.e. the difference between the full NLO result and its singular
approximation, differential in $\Tau_N$ and obtained via a
subtraction.  Here the keys to obtain stable and rapidly converging
results are the mapping that is used in the NLO calculation and the
splitting function that is used in the ``subtraction'' part. The less
the observable is modified between the $\Phi_{N+1}$ or $\Phi_{N+2}$
phase spaces and the underlying-Born $\Phi_N$ one, the more stable and
more rapidly converging the results will be.

We remark that in  order for the subtraction to be as local
as possible  \eq{finalNNLO} requires the contribution $ -  \df
  \sigma_{N,\textrm{sub.}}^{\delta\textrm{NNLO}} / (\df \Phi_N \df \Tau_N) \
P(\Phi_{N+1})$ to approximate as much as possible  the sum of the real-virtual and
the double real contributions, not only in the $\Tau_N$ variable but also in
the flavour indexes, which are omitted in the previous formulae.
This is quite nontrivial to obtain at ${\cal O}(\alpha_s^2)$, given that the subtraction term is usually obtained by the singular spectrum
which is summed over the possible flavours and partonic channels, and in order to make it differential over the flavours again one would have to devise splitting functions which carry
the correct  $\Tau_N$ and flavour dependence at ${\cal O}(\alpha_s^2)$.

The third line of \eq{finalNNLO} is just the difference between the NLO corrections of the cross-section with one extra jet  for the observable evaluated on the exact
or projected kinematics, over the entirety of the phase space. The mapping used in this projection also plays an important role in obtaining rapidly converging
results in the P2B method. The consequences of these choices are further discussed in \sec{p2b}.

In general the value of $\Tau_N^\cut$ can be freely taken and the result
is independent from this choice due to the compensating effect between the first and
the second line. It is however better to choose a value that makes those two
compensating terms roughly of the same size, in order to minimize the numerical errors in the combination.
The value of $\Tau_\delta$ must instead be chosen as small as permitted by the stability of the NLO calculation above it, in order to make the neglected inclusive power corrections as small as possible.

\section{Implementation details}
\label{sec:genimpl}
In this section we discuss the implementation of \eq{finalNNLO} into the \geneva code,
also including the fiducial power corrections below the $\Tau_\delta$ tiny cutoff.
We start by reminding the reader
that $\df \sigma_{N+1}^{\delta\textrm{NLO}} / \df \Phi_{N+1} \
\Obs(\Phi_{N+X})$ is in general not known analytically and must therefore be obtained numerically using a NLO local subtraction.
This is achieved in \geneva with the FKS subtraction method~\cite{Frixione:1995ms,Frederix:2009yq}. Omitting counterterms for ease of notation this reads
\begin{align}
\frac{\df \sigma_{N+1}^{\delta\textrm{NLO}}}{\df \Phi_{N+1}} \ \Obs(\Phi_{N+X}) & = R\!V(\Phi_{N+1})  \Obs(\Phi_{N+1}) + \int  \frac{\df \Phi_{N+2}}{\df \Phi_{N+1}}  R\!R(\Phi_{N+2})  \Obs(\Phi_{N+2})\,.
\end{align}
In \geneva, the real corrections of the NLO$_{N+1}$ calculation are separated according to the value of $\Tau_{N+1}(\Phi_{N+2})$
into different \emph{jet bins} contributions using the resolution cutoff $\Tau_{N+1}^{\rm cut}$,
depending if $\Tau_{N+1}  (\Phi_{N+2}) < \Tau_{N+1}^{\rm cut} $ or not.
In order to cover the  phase space entirely, the
nonsingular and nonprojectable contributions from the $\Phi_{N+2}$ phase space with $\Tau_{N+1}  (\Phi_{N+2}) < \Tau_{N+1}^{\rm cut} $,
which are not obtainable through a splitting of $\Phi_{N+1}$ phase space below the resolution cutoff, are included
in the $N+2$ jet bin. We indeed partition the part of the $\Phi_{N+2}$ phase space which is not removed by the process-defining cuts with a set of mutually exclusive jet-bin
theta functions $ \Theta_{N+1} (\Phi_{N+2})\,, \Theta_{N+2} (\Phi_{N+2})$, which obey $\Theta_{N+1} (\Phi_{N+2}) + \Theta_{N+2} (\Phi_{N+2}) \equiv 1$. The latter,
$\Theta_{N+2} (\Phi_{N+2})$ corresponds to
\begin{align} \label{eq:restriction_theta_np2}
    \Theta_{N+2} (\Phi_{N+2}) =  \theta\big( \Tau_{N+1} (\Phi_{N+2}) > \Tau_{N+1}^{\rm cut} \big)\ +\ \bar \Theta_{\rm flav.} (\Phi_{N+2}) \ +\ \bar \Theta_{\rm map} (\Phi_{N+2})
    \,,
\end{align}
where $\bar \Theta_{\rm flav.}$ and $\bar \Theta_{\rm map}$ represent invalid flavour and map projection of the two-closest partons, respectively.
These correspond to configurations which are nonsingular in $\Tau_{N+1}$  (despite also having $\Tau_{N+1} (\Phi_{N+2})< \Tau_{N+1}^{\rm cut}$).
A map projection can be invalid, e.g., if the map does not cover the entirety of the phase space.
The former,
$\Theta_{N+1} (\Phi_{N+2})$ is instead the complement of $\Theta_{N+2}$ in the region  which is not removed by the process-defining cuts (and not the complement of the whole  $\Phi_{N+2}$ in general)
\begin{align} \label{eq:restriction_theta_np1}
    \Theta_{N+1} (\Phi_{N+2}) =  \theta\big( \Tau_{N+1} (\Phi_{N+2}) < \Tau_{N+1}^{\rm cut} \big)\ \times\  \Theta_{\rm flav.} (\Phi_{N+2}) \ \times\ \Theta_{\rm map} (\Phi_{N+2})
    \,.
\end{align}
This is implemented through a combination of restrictions on the $\Phi_{N+1}$  phase space and on the splitting $\Phi_{N+1} \to \Phi_{N+2}$
that ensures unitarity  within the generation cuts, \emph{cfr.} \sec{gencuts}.
Expressing the NLO$_{N+1}$ corrections in terms of their actual contributions and taking also into account \eqs{restriction_theta_np2}{restriction_theta_np1},
we can rewrite \eq{finalNNLO}, after some simple manipulation, as follows
\begin{align}
\label{eq:implfiducial}
  \Obs_{\delta\textrm{NNLO}} (\Phi_N)  = &  \frac {
                                           \df
                                                 \Sigma_{N,\textrm{sub.}}^{\delta\textrm{NNLO}}}{
                                                 \df \Phi_N} \big( \Tau_N^{\textrm{cut}}\big)   \,\Obs(\Phi_{N})  \\
           &  +  \Bigg \{ \int  \frac {\df \Phi_{N+1}}{\df \Phi_{N}} \ \theta\big(\Tau_N (\Phi_{N+1}) > \Tau_\delta \big) \ \bigg [
          R\!V(\Phi_{N+1})
          \nn \\ & \qquad \qquad  + \int  \frac {\df \Phi_{N+2}}{\df \Phi_{N+1}}
          \big[ R\!R\ \Theta_{N+1} \big] (\Phi_{N+2})\  \theta\big(\Tau_N (\Phi_{N+2}) > \Tau_\delta \big) \nn\\
	& \qquad \qquad  - \frac {\df \sigma_{N,\textrm{sub.}}^{\delta\textrm{NNLO}}}{\df
         \Phi_N \df \Tau_N} \ P(\Phi_{N+1})\ \theta\big(\Tau_N (\Phi_{N+1}) <
          \Tau_N^{\textrm{cut}}\big)   \bigg]   \nn \\
     &   \qquad + \int  \frac {\df \Phi_{N+2}}{\df \Phi_{N}} \bigg[   \big[R\!R\
    \Theta_{N+1} \big] (\Phi_{N+2})\  \theta\big(\Tau_N (\widetilde \Phi_{N+1}) < \Tau_\delta \big)    \nn \\
  & \qquad \qquad + \big[R\!R\
          \Theta_{N+2} \big] (\Phi_{N+2})  \bigg] \ \theta\big(\Tau_N (\Phi_{N+2}) >
       \Tau_\delta \big)    \Bigg \}\ \Obs(\Phi_{N})  \nn\\
 &  +   \int  \frac {\df \Phi_{N+1}}{\df \Phi_{N}} \bigg [
          R\!V(\Phi_{N+1}) \big[ \Obs(\Phi_{N+1}) { - \Obs(\Phi_{N})}\big]
          \nn \\ & \qquad \qquad   + \int  \frac {\df \Phi_{N+2}}{\df \Phi_{N+1}}
          \big[ R\!R\ \Theta_{N+1} \big] (\Phi_{N+2})  \big[ \Obs(\Phi_{N+2}) { -
                   \Obs(\Phi_{N})}\big]  \bigg] \nn\\
     &  +   \int  \frac {\df \Phi_{N+2}}{\df \Phi_{N}}  \big[R\!R\
          \Theta_{N+2} \big] (\Phi_{N+2})  \big[ \Obs(\Phi_{N+2}) { - \Obs(\Phi_{N})}\big]  \nn\,.
\end{align}
This formula is better adapted to the internal structure of \geneva, in
particular separating the integration of the $\Phi_{N+2}$ phase space points
that are directly obtained from the phase space generator from the ones that
are obtained through a $\Phi_{N+1} \to \Phi_{N+2}$ splitting, which are sampled
differently.
In the fifth line of \eq{implfiducial} the $ \theta\big(\Tau_N (\widetilde \Phi_{N+1}) <
\Tau_\delta \big)$ term is evaluated on a projected  $\Phi_{N+2} \to
\widetilde \Phi_{N+1}$ point, which is in general different from the $\Phi_{N+1}$
generated point. This contribution adds back the  $\Phi_{N+2}$ points with $\theta\big(\Tau_N (\Phi_{N+2}) >
       \Tau_\delta \big) $ that
cannot be generated via the $\Phi_{N+1} \to \Phi_{N+2}$ splitting because they
fail the initial condition $\theta\big(\Tau_N (\Phi_{N+1}) >
\Tau_\delta \big) $ before the splitting. Obviously, choosing a splitting and a
projection that preserve the value of $\Tau_N$ will remove this additional contribution.
This complication is not present in the last three lines featuring the P2B subtraction because the IR  cutoff  $\Tau_\delta$ is not required there. This means that the real-virtual and double real contributions are evaluated without a lower cutoff, but they are contributing only when the difference between the observable evaluated on the exact or projected kinematics is non vanishing.

\subsection{Exclusion of fiducial power corrections}
For completeness we also present the  formula that one obtains when we do not exploit the P2B method and neglect the FPCs:
\begin{align}
  \label{eq:finalnofpc}
  \Obs_{\delta\textrm{NNLO}} (\Phi_N)  = &  \frac {
                                           \df
                                                 \Sigma_{N,\textrm{sub.}}^{\delta\textrm{NNLO}}}{
                                                 \df \Phi_N} \big( \Tau_N^{\textrm{cut}}\big)   \,\Obs(\Phi_{N})  \\
           &  +   \int  \frac {\df \Phi_{N+1}}{\df \Phi_{N}}\ \theta\big(\Tau_N (\Phi_{N+1}) > \Tau_\delta \big) \ \bigg [
          R\!V(\Phi_{N+1})  \Obs(\Phi_{N+1})
          \nn \\ & \qquad  + \int  \frac {\df \Phi_{N+2}}{\df \Phi_{N+1}}
          \big[ R\!R\ \Theta_{N+1} \big] (\Phi_{N+2}) \Obs(\Phi_{N+2})  \theta\big(\Tau_N (\Phi_{N+2}) > \Tau_\delta \big) \nn\\
	& \qquad  - \frac {\df \sigma_{N,\textrm{sub.}}^{\delta\textrm{NNLO}}}{\df
         \Phi_N \df \Tau_N} \ P(\Phi_{N+1})  \Obs(\Phi_{N}) \theta\big( \Tau_N (\Phi_{N+1}) <
          \Tau_N^{\textrm{cut}}\big)    \bigg]   \nn \\
& + \int  \frac {\df \Phi_{N+2}}{\df \Phi_{N}} \bigg[   \big[R\!R\
    \Theta_{N+1} \big] (\Phi_{N+2})\  \theta\big(\Tau_N (\widetilde \Phi_{N+1}) < \Tau_\delta \big)    \nn \\
  & \qquad + \big[R\!R\
          \Theta_{N+2} \big] (\Phi_{N+2})  \bigg] \ \Obs(\Phi_{N+2})\ \theta\big(\Tau_N (\Phi_{N+2}) >
       \Tau_\delta \big) \,.\nn
\end{align}
This is not the default, because in general one pays the price of
reintroducing observable-depend power corrections below $\Tau_\delta$,
but in in case one is interested in evaluating inclusive observables
(or those preserved by the P2B mapping) that do not suffer from FPCs,
it is better for numerical efficiency to use the simpler formula above
that avoids evaluating quantities that are eventually canceled.

\subsection{Inclusion of restrictions at the generation level }\label{sec:gencuts}
For some processes we need to impose process-defining phase space restrictions
$\thetaPSiso (\Phi_N)$ in order to have a finite cross section already at
the leading
order. We collective call this restriction as \emph{generation cuts}
even if their actual formulation could be more complicated than just a simple
cut on some variable. For example, for processes with final-state photons the
appropriate restriction might involve some photon isolation procedure.
These generation cuts affect our formulae as explained in the following.
 Additionally, we use the symbol
$\thetaProj(\widetilde{\Phi}_{N+1})$ (and
$\thetaBarProj(\widetilde{\Phi}_{N+1})$ for its complement) to indicate
this set of phase space restrictions acting on the higher dimensional
$\Phi_{N+2}$ phase space due to the cuts on the projected
configuration $\widetilde{\Phi}_{N+1}$. In practice, this means that when
a term in the cross section, evaluated at a $\Phi_{N+2}$ phase space
point, is multiplied by $\thetaProj(\widetilde{\Phi}_{N+1})$, the
$\Phi_{N+2}$ phase space point is projected onto a
$\widetilde{\Phi}_{N+1}$ point and the cuts are applied on this lower
dimensional space. If the projected configuration does not pass the
cuts, the initial $\Phi_{N+2}$ configuration is excluded.  Notice that
the separation realized by the introduction of $\thetaProj$ and
$\thetaBarProj$ is only required in our implementation of the fixed-order calculation
because of the choice of generating the $\Phi_{N+2}$ phase space starting from
$\Phi_{N+1}$ satisfying $ \thetaPSiso(\Phi_{N+1})$ instead of starting from all $\Phi_{N+1}$ points.
In order to keep the formulae simpler we continue to only consider observables that are not affected by FPCs; we will comment on how to extend these results to include FPCs in \sec{p2b}.
In this case   \eq{finalnofpc} becomes
\begin{align}
  \label{eq:isocuts1}
  \Obs_{\delta\textrm{NNLO}}&(\Phi_N)  =   \frac {
                                           \df
                                                 \Sigma_{N,\textrm{sub.}}^{\delta\textrm{NNLO}}}{
                                                 \df \Phi_N} \big( \Tau_N^{\textrm{cut}}\big) \thetaPSiso(\Phi_{N})   \,\Obs(\Phi_{N})  \\
           &  +   \int  \frac {\df \Phi_{N+1}}{\df \Phi_{N}} \ \theta\big(\Tau_N (\Phi_{N+1}) > \Tau_\delta \big) \ \bigg [
          R\!V(\Phi_{N+1}) \thetaPSiso(\Phi_{N+1})   \Obs(\Phi_{N+1})
          \nn \\ & \ \  + \int  \frac {\df \Phi_{N+2}}{\df \Phi_{N+1}}
          \big[ R\!R\ \Theta_{N+1} \thetaPSiso \big] (\Phi_{N+2}) \thetaProj(\widetilde \Phi_{N+1})  \Obs(\Phi_{N+2})  \theta\big(\Tau_N (\Phi_{N+2}) > \Tau_\delta \big) \nn\\
	& \ \ - \frac {\df \sigma_{N,\textrm{sub.}}^{\delta\textrm{NNLO}}}{\df
         \Phi_N \df \Tau_N} \ P(\Phi_{N+1}) {\thetaPSiso(\Phi_{N})} \Obs(\Phi_{N}) \theta\big( \Tau_N (\Phi_{N+1}) <
          \Tau_N^{\textrm{cut}}\big)    \bigg]   \nn \\
     &   + \int  \frac {\df \Phi_{N+2}}{\df \Phi_{N}} \bigg[   \big[R\!R\
       \Theta_{N+1} \thetaPSiso \big] (\Phi_{N+2}) \Big( \thetaBarProj (\widetilde \Phi_{N+1}) +  \theta\big(\Tau_N (\widetilde \Phi_{N+1}) < \Tau_\delta \big) \Big)    \nn\\
     &  \qquad  \qquad \quad +  \big[R\!R\
          \Theta_{N+2} \thetaPSiso \big] (\Phi_{N+2})  \bigg] \ \Obs(\Phi_{N+2}) \theta\big(\Tau_N (\Phi_{N+2}) >
       \Tau_\delta \big)\,. \nn
\end{align}
In the equation above, the $ \thetaBarProj (\widetilde \Phi_{N+1})$ in the second-to-last line correctly reinstates  the  $\Phi_{N+2}$ contributions that could not be generated via a $\Phi_{N+1} \to \Phi_{N+2}$ splitting because of the phase space restriction  $\thetaPSiso(\Phi_{N+1})$.
It is also important to notice that the $\thetaPSiso(\Phi_{N})$ multiplying the splitting function $P(\Phi_{N+1})$  should be instead a $\thetaPSiso(\Phi_{N+1})$, because that is the actual cut on the generated points. However, if one  makes that choice the relation connecting  the integral of the spectrum $\df \sigma_{N,\textrm{sub.}}^{\delta\textrm{NNLO}}$ between $\Tau_\delta$ and  $\Tau_N^{\textrm{cut}}$ and the cumulant  $\df
\Sigma_{N,\textrm{sub.}}^{\delta\textrm{NNLO}}$ above $\Tau_\delta$ is
violated, resulting a miscancellation of the subtraction term. This problem
can be avoided only if the mapping used in the splitting functions
preserves the process defining cuts, such that one can safely make the
replacement $\thetaPSiso(\Phi_{N+1})~=~\thetaPSiso(\Phi_{N})$. Factoring out the restrictions at the generation level, \eq{isocuts1} can be written as
\begingroup
  \allowdisplaybreaks
\begin{align}
   \label{eq:isocuts}
  \Obs_{\delta\textrm{NNLO}}&(\Phi_N)  =   \frac {
                                           \df
                                                 \Sigma_{N,\textrm{sub.}}^{\delta\textrm{NNLO}}}{
                                                 \df \Phi_N} \big( \Tau_N^{\textrm{cut}}\big) \thetaPSiso(\Phi_{N})   \,\Obs(\Phi_{N})  \\
           &  +   \int  \frac {\df \Phi_{N+1}}{\df \Phi_{N}} \thetaPSiso(\Phi_{N+1}) \theta\big(\Tau_N (\Phi_{N+1}) > \Tau_\delta \big) \ \bigg [
          R\!V(\Phi_{N+1})    \Obs(\Phi_{N+1})
          \nn \\ & \quad \quad  + \int  \frac {\df \Phi_{N+2}}{\df \Phi_{N+1}}
          \big[ R\!R\ \Theta_{N+1} \thetaPSiso \big] (\Phi_{N+2})   \Obs(\Phi_{N+2})  \theta\big(\Tau_N (\Phi_{N+2}) > \Tau_\delta \big) \nn\\
	& \quad \quad - \frac {\df \sigma_{N,\textrm{sub.}}^{\delta\textrm{NNLO}}}{\df
         \Phi_N \df \Tau_N} \ P(\Phi_{N+1}) \Obs(\Phi_{N}) \theta\big( \Tau_N (\Phi_{N+1}) <
          \Tau_N^{\textrm{cut}}\big)    \bigg]   \nn \\
     &   + \int  \frac {\df \Phi_{N+2}}{\df \Phi_{N}} \bigg[   \big[R\!R\
       \Theta_{N+1} \thetaPSiso \big] (\Phi_{N+2}) \Big( \thetaBarProj (\widetilde \Phi_{N+1}) +  \theta\big(\Tau_N (\widetilde \Phi_{N+1}) < \Tau_\delta \big) \Big)    \nn\\
     &  \qquad  \qquad \quad +  \big[R\!R\
          \Theta_{N+2} \thetaPSiso \big] (\Phi_{N+2})  \bigg] \ \Obs(\Phi_{N+2}) \theta\big(\Tau_N (\Phi_{N+2}) >
       \Tau_\delta \big)\,. \nn
\end{align}
\endgroup

\subsection{Usage of dynamical cuts}
When dealing with multi-scale processes, it may be advantageous to
employ dynamical generation and resolution cuts to better capture the
richer kinematic structures that could logarithmically enhance
certain phase space regions. This is also more efficient from a
numerical point of view: if the nonsingular contribution is not
particularly enhanced in a corner of the phase space, the
corresponding IR cutoff can safely be raised without paying too
hefty a price in terms of missing power corrections. The increased
IR cutoff  delivers in turn much stabler fixed-order corrections.
We express the dependence of the generation and resolution cuts on the generic phase space point though the substitutions $\Tau_\delta \to \Tau_\delta(\Phi_{N+X})$ and  $\Tau_N^{\textrm{cut}} \to \Tau_N^{\textrm{cut}}(\Phi_{N+X})$, respectively, and we write
\begin{align}
  \label{eq:dyncuts}
	\Obs_{\delta\textrm{NNLO}}& (\Phi_N)  =   \frac {
		\df
		\Sigma_{N,\textrm{sub.}}^{\delta\textrm{NNLO}}}{
		\df \Phi_N} \big( \Tau_N^{\textrm{cut}}(\Phi_N)\big) \thetaPSiso(\Phi_{N})   \,\Obs(\Phi_{N})  \\
	&  +   \int  \frac {\df \Phi_{N+1}}{\df \Phi_{N}} \bigg\{ \thetaPSiso(\Phi_{N+1})  \theta\big(\Tau_N (\Phi_{N+1}) > \Tau_\delta(\Phi_{N+1}) \big) \bigg [
	R\!V(\Phi_{N+1})    \Obs(\Phi_{N+1})
	\nn \\ & \ \  +  \int  \frac {\df \Phi_{N+2}}{\df \Phi_{N+1}}
	\big[ R\!R\  \Theta_{N+1} \thetaPSiso \big] (\Phi_{N+2})   \Obs(\Phi_{N+2})  \theta\big(\Tau_N (\Phi_{N+2}) > \Tau_\delta(\Phi_{N+2}) \big)   \nn\\
	& \ \  - \frac {\df \sigma_{N,\textrm{sub.}}^{\delta\textrm{NNLO}}}{\df
		\Phi_N \df \Tau_N} \ P(\Phi_{N+1}) \Obs(\Phi_{N}) \theta\big(  \Tau_N (\Phi_{N+1}) <
	{\Tau_N^{\textrm{cut}}(\Phi_{N+1})}\big)   \bigg ] \bigg \}   \nn \\
  &   + \int  \frac {\df \Phi_{N+2}}{\df \Phi_{N}} \bigg[   \big[R\!R\
       \Theta_{N+1} \thetaPSiso \big] (\Phi_{N+2}) \Big( \thetaBarProj (\widetilde \Phi_{N+1}) +  \theta\big(\Tau_N (\widetilde \Phi_{N+1}) < \Tau_\delta (\widetilde \Phi_{N+1})\big) \Big)    \nn\\
     &  \qquad  \qquad \quad +  \big[R\!R\
          \Theta_{N+2} \thetaPSiso \big] (\Phi_{N+2})  \bigg] \ \Obs(\Phi_{N+2}) \theta\big(\Tau_N (\Phi_{N+2}) >
       \Tau_\delta (\Phi_{N+2}) \big)\,. \nn
\end{align}
As done for \eq{isocuts}, this formula only works under the important assumption that
the mapping used in the splitting functions preserves both the process defining cuts (as before)
but now also the dynamical cuts, \emph{i.e.} $\Tau_\delta(\Phi_{N+1}) = \Tau_\delta(\Phi_{N})$ and
$\Tau_N^{\textrm{cut}}(\Phi_{N+1}) = \Tau_N^{\textrm{cut}}(\Phi_{N})$.
Note that if the splitting mapping used in the NLO$_{N+1}$ calculation does not
preserve the defining functional dependence of $\Tau_\delta(\Phi_{N+2})$, in order to capture the points that were excluded before the splitting,
the dynamic cutoff in the last line of  \eqref{eq:dyncuts} needs also to be  evaluated on the projected configuration,  $\Tau_\delta(\widetilde \Phi_{N+1})$, which can result in a different numerical value.

\subsection{Mappings for the P2B method}
\label{sec:p2b}

At this point we are ready to discuss how the fiducial power corrections can
be incorporated again, even in the presence of phase-space restrictions at the
generation level and of dynamical cuts.
Combining~\eq{implfiducial} and~\eq{dyncuts} we  obtain
\begin{align}
  \label{eq:finalP2B}
	\Obs_{\delta\textrm{NNLO}}& (\Phi_N)  =  \Bigg\{  \frac {
		\df
		\Sigma_{N,\textrm{sub.}}^{\delta\textrm{NNLO}}}{
		\df \Phi_N} \big( \Tau_N^{\textrm{cut}}(\Phi_N)\big) \thetaPSiso(\Phi_{N})   \\
	&  +   \int  \frac {\df \Phi_{N+1}}{\df \Phi_{N}} \bigg\{ \thetaPSiso(\Phi_{N+1})  \theta\big(\Tau_N (\Phi_{N+1}) > \Tau_\delta(\Phi_{N+1}) \big) \bigg [
	R\!V(\Phi_{N+1})
	\nn \\ & \qquad \qquad   +  \int  \frac {\df \Phi_{N+2}}{\df \Phi_{N+1}}
	\big[ R\!R\  \Theta_{N+1} \thetaPSiso \big] (\Phi_{N+2})   \theta\big(\Tau_N (\Phi_{N+2}) > \Tau_\delta(\Phi_{N+2}) \big)  \nn\\
	& \qquad \qquad    - \frac {\df \sigma_{N,\textrm{sub.}}^{\delta\textrm{NNLO}}}{\df
		\Phi_N \df \Tau_N} \ P(\Phi_{N+1})  \theta\big(  \Tau_N (\Phi_{N+1}) <
	{\Tau_N^{\textrm{cut}}(\Phi_{N+1})}\big) \bigg ]    \bigg \}   \nn \\
  &   + \int  \frac {\df \Phi_{N+2}}{\df \Phi_{N}} \Bigg[   \big[R\!R\
       \Theta_{N+1} \thetaPSiso \big] (\Phi_{N+2}) \Big( \thetaBarProj (\widetilde \Phi_{N+1}) +  \theta\big(\Tau_N (\widetilde \Phi_{N+1}) < \Tau_\delta (\widetilde \Phi_{N+1})\big) \Big)    \nn\\
     &  \qquad  \qquad \quad +  \big[R\!R\
          \Theta_{N+2} \thetaPSiso \big] (\Phi_{N+2})  \Bigg]  \theta\big(\Tau_N (\Phi_{N+2}) >
       \Tau_\delta (\Phi_{N+2}) \big)  \Bigg \}\ \Obs(\Phi_{N})\  \nn\\
 &  +   \int  \frac {\df \Phi_{N+1}}{\df \Phi_{N}} \thetaPSiso(\Phi_{N+1}) \bigg [
          R\!V(\Phi_{N+1}) \big[  \Obs(\Phi_{N+1}) { - \Obs(\Phi_{N})}\big]
          \nn \\ & \qquad \qquad   + \int  \frac {\df \Phi_{N+2}}{\df \Phi_{N+1}}
          \big[ R\!R\ \Theta_{N+1}  \thetaPSiso \big] (\Phi_{N+2})  \big[
                   \Obs(\Phi_{N+2}) { -
                   \Obs(\Phi_{N})}\big]  \bigg] \nn\\
     &  +   \int  \frac {\df \Phi_{N+2}}{\df \Phi_{N}} \thetaPSiso(\Phi_{N+2}) \bigg[   \big[R\!R\
       \Theta_{N+1} \big] (\Phi_{N+2}) \thetaBarProj (\widetilde
       \Phi_{N+1}) + \big[R\!R\
          \Theta_{N+2} \big] (\Phi_{N+2}) \bigg] \nn\\ & \qquad \qquad \qquad \qquad
                                                         \times \ \big[
       \Obs(\Phi_{N+2}) { -  \Obs(\Phi_{N})}\big]  \nn\,.
\end{align}

It is important to point out that in the equation above there are three
different mappings involved: \begin{enumerate*}[label=(\roman*)] \item the
  splitting mapping $\Phi_{N+1} \to \Phi_{N+2}$ \label{item:nlo}
used in the NLO calculation, \item the projection $\Phi_{N+1} \to \Phi_{N}$ used
by the splitting function in the subtraction counterterm \label{item:split}
and \item the
projections $\Phi_{N+1,2} \to \Phi_{N}$ used in evaluating the observable
that act as subtraction according to the P2B method.\label{item:p2b}
\end{enumerate*} While there is no
constraint for the mapping \ref{item:nlo}, we have already explained that the
projection used in \ref{item:split} for the splitting functions is severely
constrained to preserve
the generation restrictions. Moreover, the last projection \ref{item:p2b} of
the P2B method also needs to preserve the
generation cuts, such that one can safely write $ \thetaPSiso(\Phi_{N+2})=
\thetaPSiso(\Phi_{N+1})= \thetaPSiso(\Phi_{N})$ and derive the formula above.
This is the assumption  we have relied upon in the derivation of \eq{finalP2B},
which is satisfied by the implementations studied in this paper.
A possible alternative is to avoid the usage of generations restrictions altogether
and rely instead on a suppression factor that appropriately regulates the
divergencies. In this case one can avoid the requirements on the P2B mappings,
but the inverse of the suppression factor, which can be a very large number
near the IR limits, needs to be included in the event weight, affecting every
observable calculated from it. This procedure, while theoretically correct,
requires the cancellation between very large numbers and, as such, is
more prone to numerical inaccuracies and spikes in the final distributions.
The results we obtained during our initial tests of this alternative
approach seem to confirm this behaviour.

\section{Results and validation}
\label{sec:implementation}

In this section we discuss the numerical implementation,
presenting separately our  results for the cases
of $Z$ boson and $Z$+jet production at the LHC.
We consider the processes
\begin{align}
p p \to  \gamma^*/Z (\to \ell^+ \ell^-) (+ {\text{jet}}) +X \nonumber \, ,
\end{align}
at $\sqrt{s} = 13$ TeV.
The results are obtained in the electroweak $G_\mu$-scheme, using a complex mass scheme for the unstable internal particles.
We adopted the following values for the input parameters:
\begin{align}
m_{\text{Z}} &= 91.1876 \, \text{GeV}\,,\qquad
&\Gamma_{\text{Z}} &= 2.4952 \text{GeV}\,, \nn \\
m_{\text{W}} &= 80.379 \, \text{GeV}\,,\qquad
&\Gamma_{\text{W}} &= 2.0850 \text{GeV}\,, \nn \\
 m_{\text{t}} &= 173.1 \, \text{GeV}\,, \qquad
&G_\mu &= 1.1663787\times10^{-5}\text{GeV}^{-2}\,. \nn
\end{align}
We use $\texttt{NNPDF31\_nnlo\_as\_0118}$ as PDF
set~\cite{Ball:2017nwa}. All the matrix elements used in the
NLO calculations above $\Tau_\delta$ are taken from \openloopsTwo~\cite{Buccioni:2019sur}.

%%%%%%%%%%%%%%%%%%%%%%%%%%%%%%%%%%%%%%%%%%%%%%%%%%%%%%%%%%%%%%%%%%%%%%%%%%%%%%%%
\subsection{Results for neutral  Drell-Yan production}
\label{sec:dy}
%%%%%%%%%%%%%%%%%%%%%%%%%%%%%%%%%%%%%%%%%%%%%%%%%%%%%%%%%%%%%%%%%%%%%%%%%%%%%%%

For the  Drell-Yan  process, the  factorisation and renormalization scales are both set equal to the dilepton invariant mass,
$ \mu_R = \mu_F =  M_{\ell^+ \ell^-}$.
We impose generation cuts on the invariant mass of the lepton anti-lepton pair  such that
\begin{align}
  \thetaPSiso(\Phi_0)=\theta\big(50\, \mathrm{GeV} \leq M_{\ell^+ \ell^-} \leq 150\, \mathrm{GeV}\big)\,. \nn
\end{align}
Since the phase space mappings employed for colour-singlet production processes in \textsc{Geneva} always preserve $M_{\ell^+ \ell^-}$, for simplicity, we omit this condition from the formulas, as it factorizes uniformly across all terms.

Given that the NLO results can easily be obtained with a local subtraction and our main goal is to study the precision and numerical accuracy of our approach to calculate the NNLO corrections,
in the following we only present results for the most-computationally-demanding pure NNLO correction coefficients.
For this reason, we do not include any scale variations, which are only meaningful in the complete NNLO predictions.
However, the seven-point independent variations of $(\mu_R, \mu_F)$ have been validated along with the central predictions that we
show in this section.
Since the separation between different initial partonic channels does also depend on the factorization scale, we do not present results separated according to the initial state flavours, despite the strong cancellations that there might be between the NNLO corrections to different partonic luminosities.

We focus on the total NNLO coefficient and to highlight the dependence of them on the available phase space we show  a limited set of NNLO distributions: $\df \sigma / \df M_{\ell^+ \ell^-}^2$, $\df \sigma / \df Y_{\ell^+\ell^-}$, $\df \sigma / \df p_T^{\ell^-}$ and $ \df \sigma / \df p_T^{\ell^+}$.
For the total NNLO coefficient we also present results obtained in a fiducial region defined by the following ATLAS cuts~\cite{ATLAS:2019zci}
\begin{align}
 66\,\,\textrm{GeV}  \leq M_{\ell^+ \ell^-}  &\leq 116\,\,\mathrm{GeV}, \quad |Y_{\ell^-}|\leq 2.5, \quad \,|Y_{\ell^+}|   \leq 2.5, \\  & p^{\ell^-}_T \ge 27\,\mathrm{GeV}, \quad  p^{\ell^+}_T \ge 27\,\mathrm{GeV}\, .\nn
\end{align}

%%%%%%%%%%%%%%%%%%%%%%%%%%%%%%%%%%%%%%%%%%%%%%%%%%%%%%%%%%%%%%%%%%%%%%%%%%%%%%%%
\subsubsection{Pure Slicing and Zero-jettiness subtraction}
\label{sec:dy_slicing_subtraction}

Starting from the general subtraction formula derived in \eqref{eq:dyncuts}, we specify the expression to the case of zero-jettiness $\Tau_0$, which is a possible resolution variable for the Drell-Yan production process.
We  employ a dynamical IR cutoff which depends on $M_{\ell^+ \ell^-}$  as
\begin{align}\label{eq:tau0delta}
	\Tau_\delta(M_{\ell^+ \ell^-}) = 10^{-4} \cdot M_{\ell^+ \ell^-} ,
\end{align}
and, given the restrictions on the generated phase space, implies that $\Tau_\delta$ varies in the range $5\cdot10^{-3}\, \mathrm{GeV} < \Tau_\delta(M_{\ell^+ \ell^-}) < 1.5\cdot10^{-2}\, \mathrm{GeV}$. The upper limit where the subtraction is applied, $\Tau^{\mathrm{cut}}_0(M_{\ell^+ \ell^-})\gg\Tau_\delta(M_{\ell^+ \ell^-})$, is also chosen to be dynamical as
\begin{align}\label{eq:tau0cut}
	\Tau^{\mathrm{cut}}_0(M_{\ell^+ \ell^-}) = 10^{-2} \cdot M_{\ell^+ \ell^-} \, ,
\end{align}
so that it varies in the range $0.5\, \mathrm{GeV} < \Tau^{\mathrm{cut}}_0(M_{\ell^+ \ell^-}) < 1.5\, \mathrm{GeV}$.
We find
\begin{align}
	\label{eq:0jettiness}
	\Obs_{\delta\textrm{NNLO}} (\Phi_0)  &=   \frac {
		\df
		\Sigma_{0,\textrm{sub.}}^{\delta\textrm{NNLO}}}{
		\df \Phi_0} \big( \Tau_0^{\textrm{cut}}(M_{\ell^+ \ell^-})\big) \,\Obs(\Phi_{0})    +   \int  \frac {\df \Phi_{1}}{\df \Phi_{0}} \bigg\{ \theta\big(\Tau_0 (\Phi_{1}) > \Tau_\delta(M_{\ell^+ \ell^-}) \big) \\
	& \qquad \times \ \bigg [
	R\!V(\Phi_{1})    \Obs(\Phi_{1})
	 +  \int  \frac {\df \Phi_{2}}{\df \Phi_{1}}
	\big[ R\!R\  \Theta_{1} \big] (\Phi_{2})   \Obs(\Phi_{2})  \theta\big(\Tau_0 (\Phi_{2}) > \Tau_\delta(M_{\ell^+ \ell^-}) \big)   \nn\\
	& \qquad  \qquad  - \frac {\df \sigma_{0,\textrm{sub.}}^{\delta\textrm{NNLO}}}{\df
		\Phi_0 \df \Tau_0} \ P(\Phi_{1}) \Obs(\Phi_{0}) \theta\big(  \Tau_0 (\Phi_{1}) <
	{\Tau_0^{\textrm{cut}}(M_{\ell^+ \ell^-})}\big)   \bigg ] \bigg \}   \nn \\
	&   + \int  \frac {\df \Phi_{2}}{\df \Phi_{0}} \bigg[   \big[R\!R\
	\Theta_{1}  \big] (\Phi_{2}) \theta\big(\Tau_0 (\widetilde \Phi_{1}) < \Tau_\delta (M_{\ell^+ \ell^-})\big) +  \big[R\!R\
	\Theta_{2}  \big] (\Phi_{2})  \bigg] \nn\\
	&  \qquad  \qquad \qquad \times   \, \Obs(\Phi_{2})\, \theta\big(\Tau_0 (\Phi_{2}) >
	\Tau_\delta (M_{\ell^+ \ell^-}) \big)\,. \nn
\end{align}
In the formula above, we set
\begin{align}
  \label{eq:leadpower}
\frac{\df\Sigma_{0,\textrm{sub.}}^{\delta\textrm{NNLO}}}{
\df \Phi_0} \big( \Tau_0^{\textrm{cut}}\big)  = \frac{\df\Sigma_{0}^{\textrm{NNLL}^\prime}}{
\df \Phi_0}\big( \Tau_0^{\textrm{cut}}\big) \bigg|_{\mathcal{O}(\alpha^2_s)}  & \qquad \textrm{and}\qquad
\frac{\df \sigma_{0,\textrm{sub.}}^{\delta\textrm{NNLO}}}{\df
\Phi_0 \df \Tau_0}  = \frac{\df\Sigma_{0}^{\textrm{NNLL}^\prime}}{
\df \Phi_0\df \Tau_0} \bigg|_{\mathcal{O}(\alpha^2_s)}\, ,
\end{align}
where the LP NNLL$^\prime \ \Tau_0$ results for the resummed-expanded cumulant and spectrum\footnote{We remark that the  $\ord {\as^2}$  expansion of the NNLL$^\prime$ and the N$^3$LL resummed results are identical for both the cumulant and the spectrum, so we can equivalently use either one. These are also called the singular NNLO cross section cumulant and spectrum, respectively.} on the right hand side were evaluated in Refs.~\cite{Alioli:2015toa,Alioli:2023har}.
At LP accuracy, it can be shown that the resummed-expanded cumulant and spectrum in $\Tau_0$ factorize, in the limit $\Tau^{\mathrm{cut}}_0\ll M_{\ell^+ \ell^-}$ or $\Tau_0\ll M_{\ell^+ \ell^-}$, as a product of a hard function times the convolution of two beam functions and a soft function \cite{Stewart:2009yx}.
In addition, note that in \eq{0jettiness} there is not an explicit $\thetaBarProj$ contribution since the generation cuts on $M_{\ell^+ \ell^-}$ are always preserved by the mappings. The second to last line is included to add back the points which satisfy the conditions $\theta(\Tau_0(\Phi_2)>\Tau_\delta)\,\theta(\Tau_0(\widetilde \Phi_1)<\Tau_\delta)$ and could not be generated via the FKS $\Phi_1\to \Phi_2$ splitting, which is employed in this calculation. The splitting functions $P(\Phi_1)$ also utilize a FKS mapping for the $\Phi_1\to \Phi_0$ projections, hence they also preserve $M_{\ell^+ \ell^-}$.

The slicing formula can be easily recovered from \eq{0jettiness} as a special case by setting $\Tau^{\mathrm{cut}}_0(M_{\ell^+ \ell^-})=\Tau_\delta(M_{\ell^+ \ell^-})$ as defined in \eq{tau0delta}. From this follows that the cumulant in the first line of  \eq{0jettiness} is evaluated at $\Tau_\delta(M_{\ell^+ \ell^-})$ and the subtraction term in the fourth line of \eq{0jettiness} effectively disappear from the equation.

In the first column of \Tab{dyxs} we compare the $\mathcal{O}(\alpha^2_s)$ corrections to the total Drell-Yan cross section coefficients obtained via the pure slicing method and with the nonlocal subtraction.
We first notice how these results, summed over all the partonic channels and integrated over the available phase space, result in corrections which are negative and $\leq 1\%$ of the total NLO cross section. We also find that they are consistent within integration errors, and under equivalent conditions of running time and machine type, the numerical integration error of the subtraction method is slightly reduced compared to the pure slicing implementation.

We then present the results for the cross section within the ATLAS fiducial region in the second column. Also in this case we observe a marginal reduction in the error when performing a nonlocal subtraction, despite the final result still being affected by sizable fiducial power corrections (see \sec{z_fpc}).

\begin{figure*}[ht!]
  \centering
  \begin{subfigure}[b]{\rescaletwoplots}
    \includegraphics[width=\textwidth]{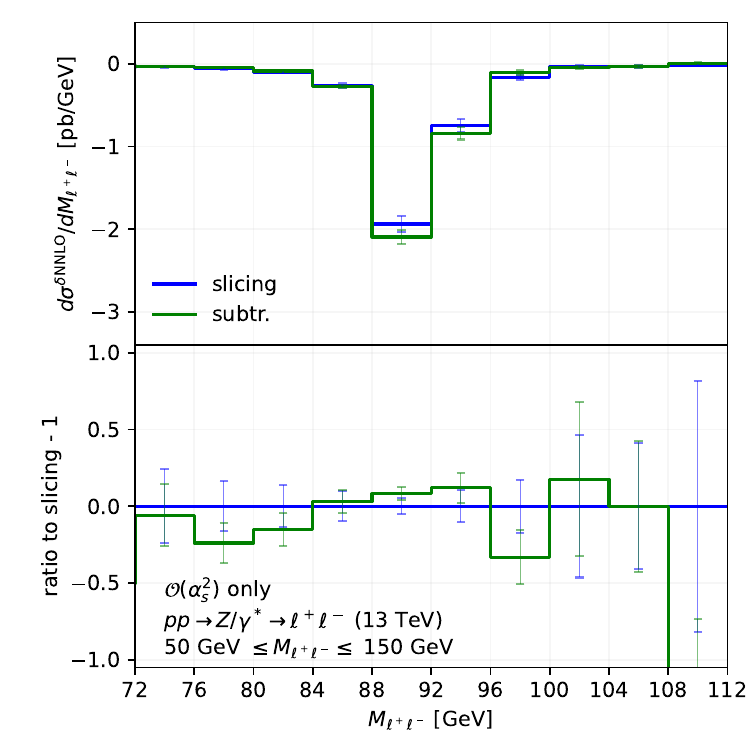}%
  \end{subfigure}
  \begin{subfigure}[b]{\rescaletwoplots}
    \includegraphics[width=\textwidth]{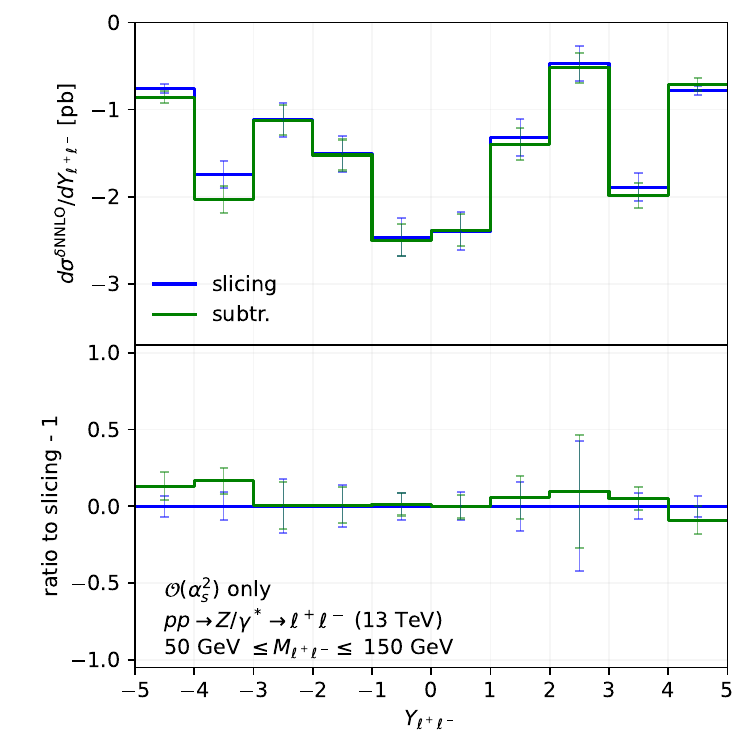}%
  \end{subfigure}
  \begin{subfigure}[b]{\rescaletwoplots}
    \includegraphics[width=\textwidth]{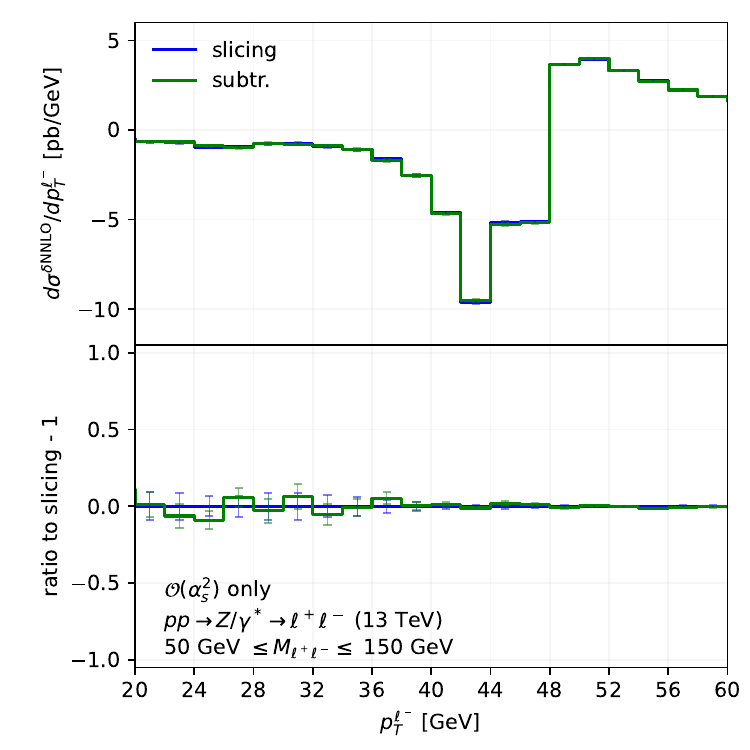}%
  \end{subfigure}
  \begin{subfigure}[b]{\rescaletwoplots}
    \includegraphics[width=\textwidth]{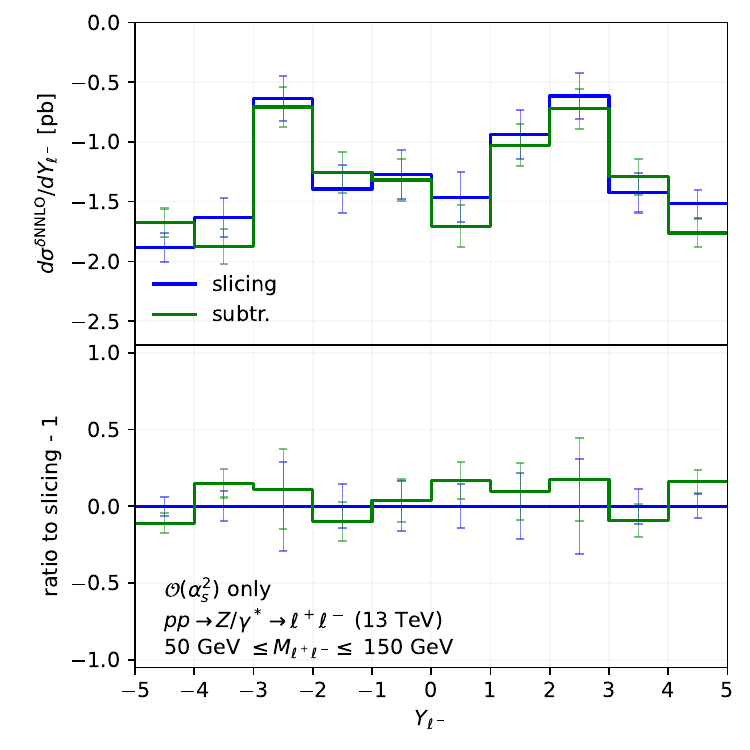}%
  \end{subfigure}
   \caption{Comparison of the Drell-Yan $\ord{\as^2}$ corrections for a set of differential distributions between the slicing and subtraction approaches.}
  \label{fig:AAA}
\end{figure*}

The comparison of the NNLO differential distributions is instead shown in \Fig{AAA}, for the  invariant mass and rapidity of the dilepton system in the upper panels and for the transverse momentum\footnote{The distribution of the transverse momentum of any lepton coming from the $Z$ decay is NNLO only below the Jacobian peak, $p_T^{\ell^\pm} \leq M_{\ell^+ \ell^-} / 2$, because the double virtual contributions can only be present there.} and rapidity of the negatively charged lepton coming from the $Z$ decay in the lower panels.  As for the more inclusive results, we observe a complete agreement and a mild reduction of the statistical uncertainties when performing the nonlocal subtraction.

\setlength{\tabcolsep}{10pt}
\begin{table}[t]
	\centering
	\begin{tabular}{|c|c|c|}
                \hline
		 Method    & ${\ord {\as^2}} $~corr. [pb]  &  ${\ord {\as^2}} $~corr. with ATLAS fid. cuts [pb] \\
		\hline
          Slicing  & $-14.45\pm 0.56$   &  $-10.13 \pm 0.33$  \\
          \hline
		Subtraction & $-15.03 \pm 0.49$   &  $-10.61 \pm 0.28$ \\
		\hline
	\end{tabular}
	\caption{Comparison of NNLO corrections to the Drell-Yan cross
          sections between the pure slicing and the nonlocal
          subtraction implementations. The reported uncertainty refers
          exclusively to the error from numerical integration.}
	\label{tab:dyxs}
\end{table}

%%%%%%%%%%%%%%%%%%%%%%%%%%%%%%%%%%%%%%%%%%%%%%%%%%%%%%%%%%%%%%%%%%%%%%%%%%%%%%%%
\subsubsection{Adding subleading inclusive power corrections}
\label{sec:dy_nlp}
%%%%%%%%%%%%%%%%%%%%%%%%%%%%%%%%%%%%%%%%%%%%%%%%%%%%%%%%%%%%%%%%%%%%%%%%%%%%%%%

Our choice of using the LP term in the expansion
of the full cross section as $\Tau_0 \to 0$ when defining
the `subtraction' cross section for $\Tau_0 \leq \Tau_\delta$
makes our predictions subject to an error, $\Delta\sigma$, that
can be inferred from the corrections that the $\Tau_0$  factorisation
theorem receives and follows a power-like trend $\Delta\sigma\sim \ord{\Tau_\delta/M_{\ell^+ \ell^-}}$~\cite{Stewart:2010qs, Stewart:2010tn}.
This is well appreciated and has been previously investigated in the literature~\cite{Gaunt:2015pea, Campbell:2019dru, Campbell:2024hjq},
leading to their calculation at NLO~\cite{Boughezal:2016zws, Ebert:2018lzn, Boughezal:2018mvf}, NNLO~\cite{Boughezal:2016zws, Moult:2016fqy, Moult:2017jsg}
and, more recently, at N$^3$LO~\cite{Vita:2024ypr}.
In these works, it was shown that despite the overall $\Tau_0$ power suppression of $\Delta\sigma$, the presence of logarithmic terms on the resolution variable necessitates extremely small values of $\Tau_\delta$ in order to meet a sensible precision goal in theoretical predictions.
In turn, this is a non-trivial requirement for the calculation that involves
(at least) one resolved emission, since the corresponding matrix elements have to be evaluated close to the IR-singular limits of QCD.
A way to mitigate this issue and systematically improve the subtractions, is to include (some of) these power suppressed terms as part of the `subtraction' cross section. This allows for higher values of $\Tau_\delta$ while keeping the same accuracy goal and reducing the numerical integration error or, as a corollary, for an improved accuracy goal while keeping $\Tau_\delta$ at the the same value.

For a practical implementation in \geneva, the $\Tau_0$ differential results of \refcite{Moult:2016fqy} for the leading-log (LL) NLP partonic
coefficient are included in the  subtracted spectrum
\begin{align} \label{eq:nnlo_nlp_spectrum} & \frac { \df
    \sigma_{0,\textrm{sub.}}^{\delta\textrm{NNLO}}}{ \df \Phi_0 \df \Tau_0}   =
  \frac{\df\sigma_{0}^{\textrm{NNLL}^\prime}}{ \df \Phi_0 \df \Tau_0 }  \bigg|_{\mathcal{O}(\alpha^2_s)}   +  \frac{1}{M_{\ell^+ \ell^-}}
  \sum _\kappa \left[  C^{(2,2)}_{\kappa,\,3} \ln^3
    \left(\frac{\Tau_0}{M_{\ell^+ \ell^-}} \right) \right] \,,
\end{align}
 where we denote the relevant
partonic channels by $\kappa = \{q\bar{q}, qg, \ldots\}$ and  the  LL NLP coefficients
$C^{(2, 2)}_{\kappa,\,3}$ are obtained by the partonic ones  in eqs.~(36)
and (37) of \refcite{Moult:2016fqy} after the convolution with the PDFs.

At this point we can  distinguish two possible ways of including the NLP corrections in the cumulant.
In the first approach, which we refer to as `spectrum log-counting', we use as `subtraction' term the LP cross section
together with the LL NLP coefficients as given in \eq{nnlo_nlp_spectrum}.
After integrating the spectrum of \eq{nnlo_nlp_spectrum} up to  $\Tau_0^\cut$  we get
%%%
\begin{align} \label{eq:nnlo_nlp_cumulant} & \frac { \df
    \Sigma_{0,\textrm{sub.}}^{\delta\textrm{NNLO}}}{ \df \Phi_0} \big(
  \Tau_0^{\textrm{cut}}\big)  =
  \frac{\df\Sigma_{0}^{\textrm{NNLL}^\prime}}{ \df \Phi_0} \big(
  \Tau_0^{\textrm{cut}}\big) \bigg|_{\mathcal{O}(\alpha^2_s)}   +  \frac{\Tau_0^{\textrm{cut}}}{M_{\ell^+ \ell^-}}
  \sum _\kappa \left[  \sum _{m=0}^3 A^{(2,2)}_{\kappa,\,m} \ln^m
    \left(\frac{\Tau_0^{\textrm{cut}}}{M_{\ell^+ \ell^-}} \right)  \right] \,,
\end{align} with
$A^{(2, 2)}_{\kappa,\,0} = - A^{(2, 2)}_{\kappa,\,1} = 2 A^{(2,
  2)}_{\kappa,\,2} = -6 A^{(2, 2)}_{\kappa,\,3} = -6 C^{(2,
  2)}_{\kappa,\,3}$.
In spectrum log-counting all the terms above must be retained such that the dependence on $\Tau_0^\cut/M_{\ell^+ \ell^-}$ properly cancels between the differential subtraction and the cumulant cross section.
In the second approach, which we refer to as `cumulant log-counting', we perform the differential subtractions between
$\Tau_\delta \leq \Tau_0 \leq \Tau^\cut_0$ using just the LP cross section as `subtraction' term
and only include in the cumulant  the highest-power logarithm of \eq{nnlo_nlp_cumulant}.
This practically means that only the term proportional to $A^{(2,2)}_{\kappa,\,3}$
is taken into account whereas $A^{(2,2)}_{\kappa,\,m}, \, m = 0,1,2$ are neglected.
This choice, albeit unfounded as it may be regarded, is no more arbitrary than keeping all terms in \eq{nnlo_nlp_cumulant} because they are anyway only obtained by the integration of the LL NLP coefficient and therefore they are missing contributions from the subleading logarithmic NLP terms in the spectrum. For this reason, it is not clear \emph{a priori} whether including (or not) the $A^{(2,2)}_{\kappa,\,m}, \, m = 0,1,2$ terms only coming from the integration of the LL NLP spectrum coefficient does improve the theoretical predictions\footnote{As far as we understand, the cumulant log-counting in a pure slicing calculation is the choice adopted by MCFM in \refscite{Campbell:2017aul, Campbell:2019dru}}.
Despite the cumulant log-counting might seem the safest choice, due to the fact that the highest logarithm does not receive contributions from the integration of unknown subleading logarithmic NLP terms, one should still be careful because only including the $A^{(2,2)}_{\kappa,\,3}$ term can miss sizable numerical cancellations with the unknown NLP subleading logarithmic terms.
Therefore, we investigate both possibilities, in order to have  a rough estimate on the size of the cancellations between the integral of NLP logarithmic terms.

\setlength{\tabcolsep}{10pt}
\begin{table}[ht!]
\centering
\begin{tabular}{|c|c|c|}
\hline
Subtraction with     & ${\ord {\as^2}} $~corr. [pb]  &  ${\ord {\as^2}} $~corr. with ATLAS fid. cuts [pb] \\
\hline
spectrum log-counting & $-13.64 \pm 0.36$   &  $-10.11 \pm 0.20$ \\
\hline
cumulant log-counting & $-14.66 \pm 0.51$   &  $-10.41 \pm 0.29$ \\
  \hline
  	\hline
  \nnlojet & $ -13.40  \pm 0.01  $  & $  -1.63  \pm 0.35 $  \\
  \hline

\end{tabular}
\caption{Comparison of the Drell-Yan $\ord{\as^2}$ total cross section coefficient using two different approaches for the
inclusion of LL NLP inclusive power corrections. The reported uncertainty refers exclusively to the error from numerical integration.}
\label{tab:dy_total_xsec_nlp}
\end{table}

\begin{figure*}[ht!]
  \centering
  \begin{subfigure}[b]{\rescaletwoplots}
    \includegraphics[width=\textwidth]{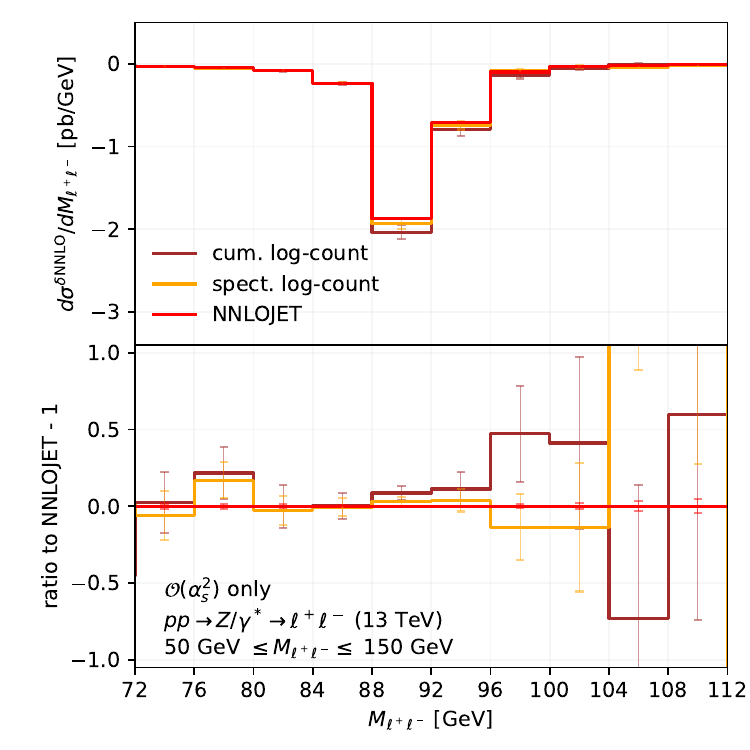}%
  \end{subfigure}
  \begin{subfigure}[b]{\rescaletwoplots}
    \includegraphics[width=\textwidth]{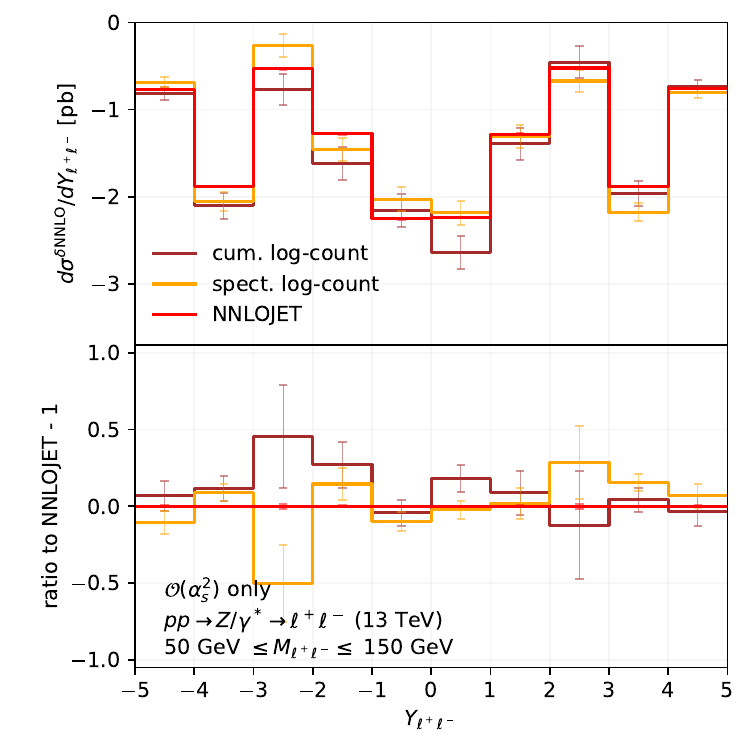}%
  \end{subfigure}
  \begin{subfigure}[b]{\rescaletwoplots}
    \includegraphics[width=\textwidth]{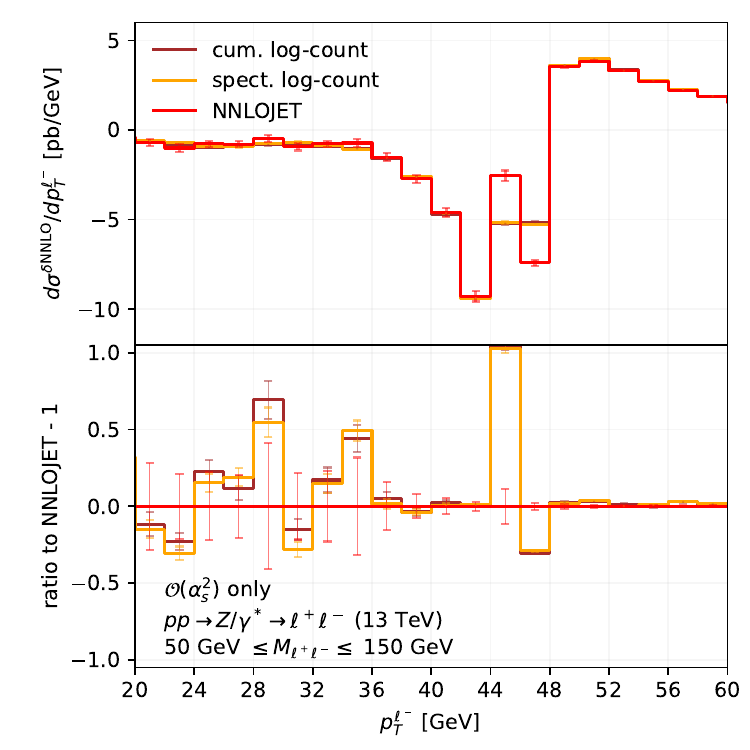}%
  \end{subfigure}
  \begin{subfigure}[b]{\rescaletwoplots}
    \includegraphics[width=\textwidth]{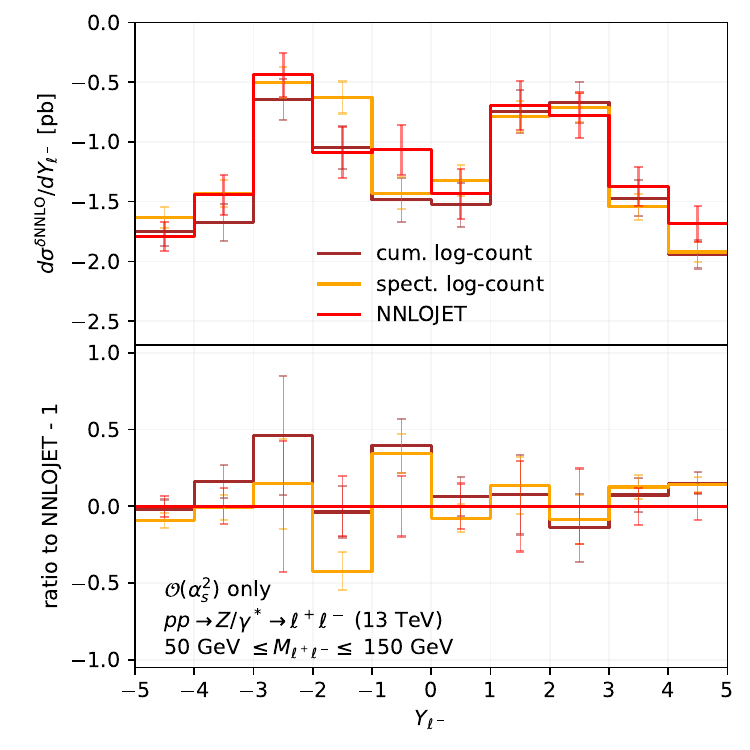}%
  \end{subfigure}
  \caption{Comparison of the Drell-Yan $\ord{\as^2}$ corrections for a set of differential distributions between \nnlojet and the two LL NLP counting schemes implementations.}
  \label{fig:nlp_cumulant_spectrum_cmp}
\end{figure*}

In \Tab{dy_total_xsec_nlp} we compare the $\mathcal{O}(\alpha^2_s)$ corrections to the total rates,  obtained with the cumulant and spectrum logarithmic countings, for both inclusive and fiducial phase space regions.
We find again that these results are consistent within integration errors, and under equivalent conditions of running time and machine type. The numerical integration errors are also similar, although the results obtained with the spectrum log-counting looks marginally better. This is a consequence of including the LL NLP terms in the subtraction of the spectrum.
In any case, we observe that for the small choice of the IR cutoff $\Tau_\delta$ used for these predictions, the effect of including the LL NLP corrections for both logarithmic countings is anyway rather small.
In the last row of the table, we report the corresponding predictions from \nnlojet.
We start by noticing how the
local NNLO subtraction in \nnlojet is particularly efficient in the predictions of the inclusive corrections. This is a peculiarity of applying the antenna subtraction to the Drell-Yan process, where the subtraction term is very similar to the Drell-Yan matrix element. Already looking at the corrections in the fiducial regions we see how the statistical errors increase.
Our results for the corrections in the ATLAS fiducial region are instead in disagreement with \nnlojet{}. This is expected since this observable is indeed strongly sensitive to FPCs which are are not yet included in the results of \Tab{dy_total_xsec_nlp}.

In \Fig{nlp_cumulant_spectrum_cmp} we show predictions for the $\ord{\as^2}$ coefficient  of a set of differential distributions for the two LL NLP counting schemes and we find very good agreement between them. Similarly to the total cross section coefficients, we also notice that the results obtained with the spectrum counting scheme show smaller statistical integration errors across all distributions.
The statistical integration error of \nnlojet is very small for the $M_{\ell^+ \ell^-}$ and $Y_{\ell^+ \ell^-}$ inclusive distributions while it sensibly increases for $p^{\ell^{-}}_T$ and $Y_{\ell^-}$.
As explained above, this is a feature of the local antenna subtraction method for this particular process.
The agreement with \nnlojet is also very good and we observe an expected difference with our calculations for the $p^{\ell^{-}}_T$ distribution in the region $44\,\mathrm{GeV} \leq p^{\ell^{-}}_T \leq 48\, \mathrm{GeV}$. Also this observable is indeed very sensitive to FPCs, which are are not yet included in our predictions. This plot  highlights the importance of their inclusion, which will be discussed in the next section (compare for example to the corresponding plot in \Fig{nlp_cumulant_spectrum_fpc_cmp}). One however should always keep in mind that any fixed-order prediction is intrinsically problematic in describing the region close to the Jacobian peak, $M_{\ell^+ \ell^-}/2$, for this distribution, because  Sudakov-shoulder logarithms become important and only after their resummation one obtains physically-sensible results \cite{Catani:1997xc, Ebert:2019zkb, Ebert:2020dfc}.

%%%%%%%%%%%%%%%%%%%%%%%%%%%%%%%%%%%%%%%%%%%%%%%%%%%%%%%%%%%%%%%%%%%%%%%%%%%%%%%%
\subsubsection{Inclusion of FPC corrections}
\label{sec:z_fpc}
%%%%%%%%%%%%%%%%%%%%%%%%%%%%%%%%%%%%%%%%%%%%%%%%%%%%%%%%%%%%%%%%%%%%%%%%%%%%%%%

Lastly, we include the fiducial power corrections, specifying \eq{finalP2B} for
zero-jettiness subtraction and using the FKS mapping recursively for the $\Phi_2 \to \Phi_1 \to \Phi_0 $ projections
which are needed for the evaluation of the subtracted observable according to the P2B method.

\setlength{\tabcolsep}{10pt}
\begin{table}[htp!]
	\centering
	\begin{tabular}{|c|c|c|}
		\hline
		 \!\!\!Subtraction with FPC and\!\! &\! \!\!${\ord {\as^2}} $~corr. [pb]\!\!  &\!\!\! ${\ord {\as^2}} $~corr.\! with ATLAS fid. cuts [pb]\!\! \\
		\hline
		spectrum log-counting & $ -13.85 \pm 0.32 $   &  $ -1.14 \pm 0.18 $ \\
		\hline
		cumulant log-counting & $ -13.21 \pm 0.33 $   &  $ -0.84 \pm 0.19$ \\
		\hline
		\hline
		\nnlojet & $ -13.40  \pm 0.01  $  & $  -1.63  \pm 0.35 $  \\
		\hline
	\end{tabular}
	\caption{Comparison with \nnlojet{} of the Drell-Yan $\ord{\as^2}$ total cross section coefficient after including FPC corrections using the two different approaches for the inclusion of LL NLP inclusive power corrections in \textsc{Geneva}. The reported uncertainty refers exclusively to the error from numerical integration.}
	\label{tab:dy_total_xsec_fpc}
\end{table}

\begin{figure*}[ht!]
  \centering
  \begin{subfigure}[b]{\rescaletwoplots}
    \includegraphics[width=\textwidth]{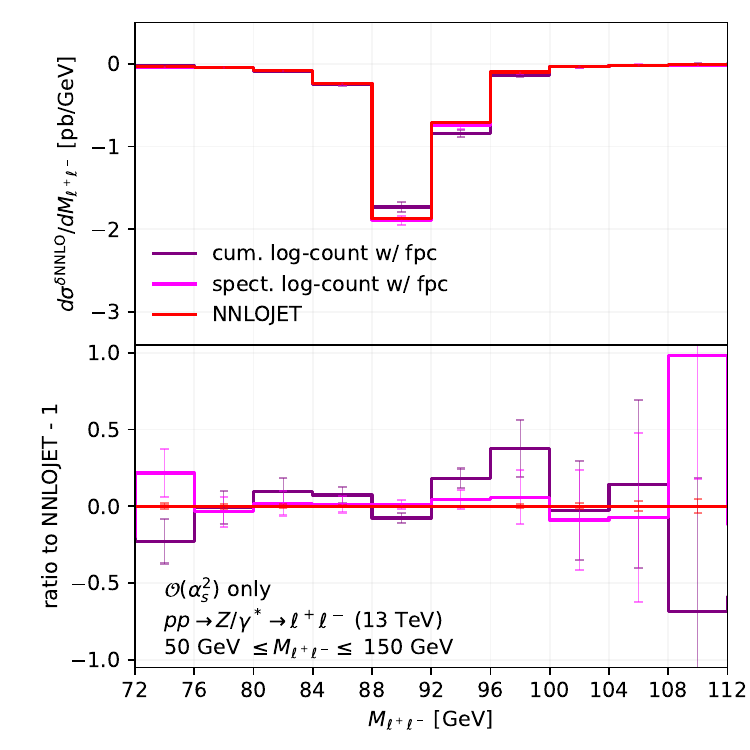}%
  \end{subfigure}
  \begin{subfigure}[b]{\rescaletwoplots}
    \includegraphics[width=\textwidth]{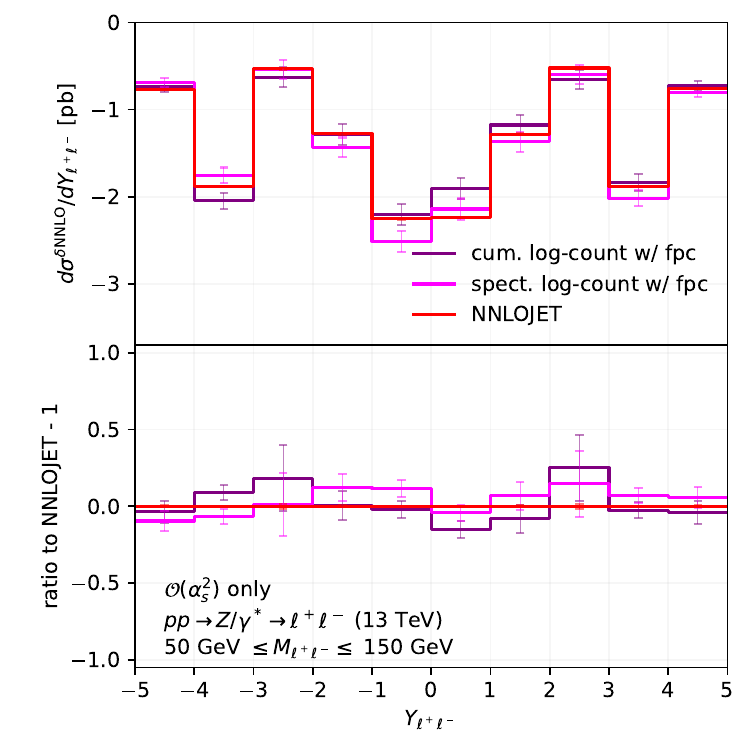}%
  \end{subfigure}
  \begin{subfigure}[b]{\rescaletwoplots}
    \includegraphics[width=\textwidth]{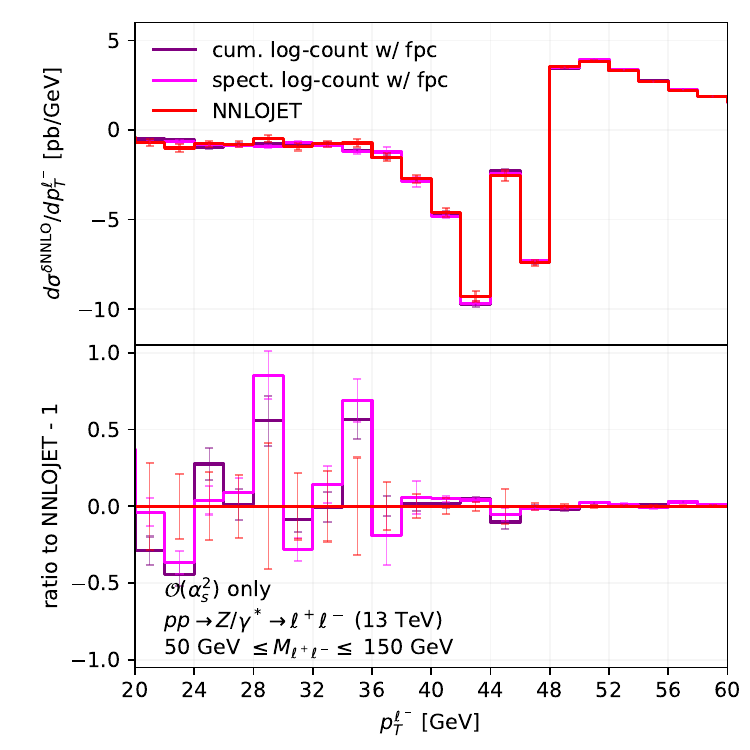}%
  \end{subfigure}
  \begin{subfigure}[b]{\rescaletwoplots}
    \includegraphics[width=\textwidth]{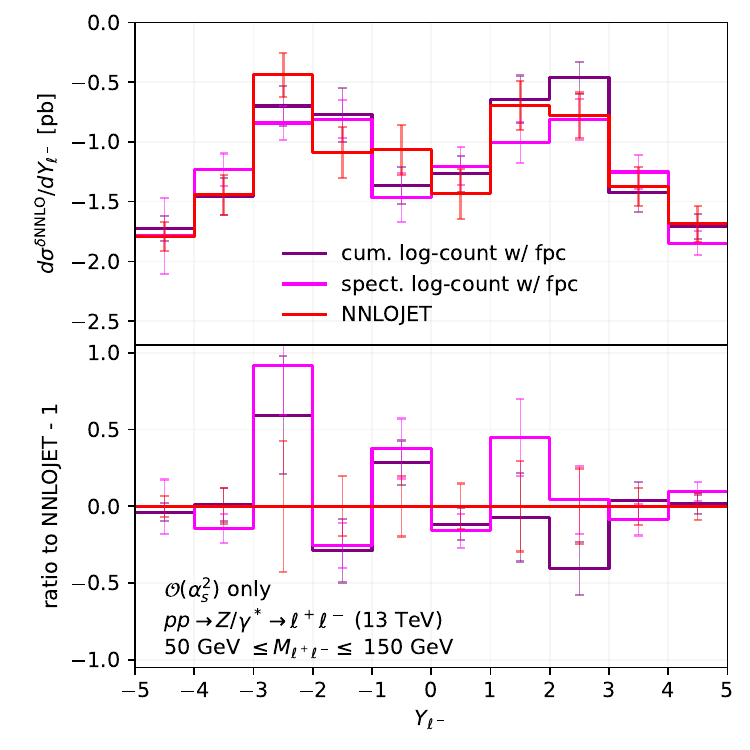}%
  \end{subfigure}
  \caption{Comparison of the Drell-Yan $\ord{\as^2}$ corrections for a set of differential distributions between \nnlojet and the two LL NLP counting schemes implementations with the inclusion of FPCs.}
  \label{fig:nlp_cumulant_spectrum_fpc_cmp}
\end{figure*}

We find
\begingroup
  \allowdisplaybreaks
\begin{align}
	\Obs_{\delta\textrm{NNLO}}& (\Phi_0)  =  \Bigg\{   \frac {
		\df
		\Sigma_{0,\textrm{sub.}}^{\delta\textrm{NNLO}}}{
		\df \Phi_0} \big( \Tau_0^{\textrm{cut}}(M_{\ell^+ \ell^-})\big)  +  \int  \frac {\df \Phi_{1}}{\df \Phi_{0}} \bigg\{ \theta\big(\Tau_0 (\Phi_{1}) > \Tau_\delta(M_{\ell^+ \ell^-}) \big) \nn \\
	& \qquad \qquad \times  \bigg [	R\!V(\Phi_{1})  +  \int  \frac {\df \Phi_{2}}{\df \Phi_{1}}
	\big[ R\!R\  \Theta_{1} \big] (\Phi_{2})   \theta\big(\Tau_0 (\Phi_{2}) > \Tau_\delta(M_{\ell^+ \ell^-}) \big)  \nn\\
	& \qquad \qquad \qquad  - \frac {\df \sigma_{0,\textrm{sub.}}^{\delta\textrm{NNLO}}}{\df
		\Phi_0 \df \Tau_0} \ P(\Phi_{1})  \theta\big(  \Tau_0 (\Phi_{1}) <
	{\Tau_0^{\textrm{cut}}(M_{\ell^+ \ell^-})}\big)  \bigg ]  \bigg \}   \nn  \\
  & \qquad  \qquad + \int  \frac {\df \Phi_{2}}{\df \Phi_{0}} \Bigg[   \big[R\!R\
       \Theta_{1} \big] (\Phi_{2}) \theta\big(\Tau_0 (\widetilde \Phi_{1}) < \Tau_\delta (M_{\ell^+ \ell^-}) \big) \\
     &  \qquad  \qquad \qquad \qquad  +  \big[R\!R\
          \Theta_{2} \big] (\Phi_{2})  \Bigg]  \theta\big(\Tau_0 (\Phi_{2}) >
       \Tau_\delta (M_{\ell^+ \ell^-}) \big)\   \Bigg \} \Obs(\Phi_{0})\  \nn\\
 &  +   \int  \frac {\df \Phi_{1}}{\df \Phi_{0}}  \bigg [
          R\!V(\Phi_{1}) \big[  \Obs(\Phi_{1}) { - \Obs(\Phi_{0})}\big]
            + \int  \frac {\df \Phi_{2}}{\df \Phi_{1}}
          \big[ R\!R\ \Theta_{1}  \big] (\Phi_{2})  \big[
                   \Obs(\Phi_{2}) { -
                   \Obs(\Phi_{0})}\big]  \bigg] \nn\\
     &  +   \int  \frac {\df \Phi_{2}}{\df \Phi_{0}} \bigg[  \big[ R\!R\ \Theta_{2}  \big] (\Phi_{2}) \big[
                                                           \Obs(\Phi_{2}) { -  \Obs(\Phi_{0})}\big] \bigg] \nn\,.
\end{align}
\endgroup

In \Tab{dy_total_xsec_fpc} we compare the $\mathcal{O}(\alpha^2_s)$ corrections to the total production cross section corrections obtained with the cumulant and spectrum logarithmic countings, also including FPCs, for both inclusive and fiducial phase space regions.
The central values for the inclusive corrections are not sensitive to the inclusion of FPCs and indeed they are compatible with the results shown in \Tab{dy_total_xsec_nlp}.
The inclusion of FPCs has instead a fundamental impact on the fiducial cross sections results computed using both counting schemes: it reduces the size of $\mathcal{O}(\alpha^2_s)$ corrections giving compatible results. Given the larger uncertainty reported by \nnlojet for this quantity, all these predictions are in agreement, resolving the tension found in \Tab{dy_total_xsec_nlp}.

In \Fig{nlp_cumulant_spectrum_fpc_cmp} we show predictions for the $\ord{\as^2}$ coefficient  of a set of differential distributions including FPCs. Overall, we observe the same behaviour as in \Fig{nlp_cumulant_spectrum_cmp} for more inclusive quantities, with the notable exception of the $p^{\ell^{-}}_T$ around the Jacobian peak, which is now in complete agreement with \nnlojet. This is an indication that our implementation correctly captures the fiducial effects.

%%%%%%%%%%%%%%%%%%%%%%%%%%%%%%%%%%%%%%%%%%%%%%%%%%%%%%%%%%%%%%%%%%%%%%%%%%%%%%%%
\subsection{Results for neutral Drell-Yan plus one jet production}
\label{sec:zjet}
%%%%%%%%%%%%%%%%%%%%%%%%%%%%%%%%%%%%%%%%%%%%%%%%%%%%%%%%%%%%%%%%%%%%%%%%%%%%%%%
Moving on to the more complicated case with an extra jet,  we now consider the process
\begin{align}
p p \to  \gamma^*/Z (\to \ell^+ \ell^-) + {\text{jet}}+X \nonumber \,.
\end{align}
As for the Drell-Yan case presented in \sec{dy}, we  present results only for the pure NNLO coefficients, which means that
all the results presented in this section are  at absolute  $\ord{\as^3}$.
We do not separate the results according to the initial state flavours neither we include any scale variation, which can only be correctly accounted for in the complete NNLO predictions.

The presence of one jet already at the leading order requires the introduction of a \emph{defining cut} in order to avoid a divergent cross-section. Several choices are possible, as long as the IR region when the jet becomes soft and/or collinear to the beam is avoided. For the present study we have limited ourselves to cutting on zero-jettiness, $\Tau_0$ or on the transverse momentum $q_T$ of the $Z$ boson. Both quantities vanish when there are no QCD emissions, making them suitable defining cuts,  but, due to its vectorial nature, $q_T$  can be zero even in the presence of two back-to-back jets, irrespective of their energy. Because of this property, the phase space regions carved out by cutting on  $\Tau_0$ or $q_T$ are intrinsically different, which gives us a good handle on exploring the dependence of our results on the defining cuts.

We therefore  show results for the $\ord{\as^3}$  cross section corrections, employing different sets of measurement cuts $\sigma^{\delta\NNLO} (\Tau_0>w)$ and $\sigma^{\delta \NNLO} (q_T>w)$ for $w=1,10,50$ and $100$~GeV.
Similarly, we only focus on a limited set of NNLO differential distributions, namely the zero-jettiness $\df \sigma / \df \Tau_0$ and the transverse momentum  $\df \sigma / \df q_T$ distributions, as well as some distributions of the leptonic decay products of the vector boson with a cut on $\Tau_0$ or $q_T$.

At variance with the Drell-Yan production case, now there are potentially more hard scales present in this process.
A suitable choice for the factorisation and renormalization scales is
%%%
\begin{align}
\mu_R = \mu_F = \mu_\FO =  m_{T} \equiv \sqrt{ M^2_{\ell^+ \ell^-} + q_T^2}
\,,\end{align}
%%%
where $M_{\ell^+ \ell^-}$ is the invariant mass and $q_T$ is the transverse momentum of the colour-singlet system.
This variable indeed captures the dynamic of the the recoil against a hard jet while providing a nonzero value even when the
QCD emissions become soft or collinear. In general however, any choice of a single scale in a multi-scale problem will inevitably
generate potentially large logarithms of different scale ratios.
When deciding the value for the IR resolution cutoff $\Tau_\delta$ adopted in the calculation we have to factor in this consideration. We will choose a dynamical cut that balances the accuracy of the calculation by limiting the potentially large scale ratios while, at the same time, avoid a resolution cut too  small which will result in an unstable cancellation between terms that are numerically large.

%%%%%%%%%%%%%%%%%%%%%%%%%%%%%%%%%%%%%%%%%%%%%%%%%%%%%%%%%%%%%%%%%%%%%%%%%%%%%%%%
\subsubsection{Slicing vs subtraction}
\label{sec:zj_slicing_subtraction}
%%%%%%%%%%%%%%%%%%%%%%%%%%%%%%%%%%%%%%%%%%%%%%%%%%%%%%%%%%%%%%%%%%%%%%%%%%%%%%%

We first implement a simpler slicing approach, using the
absolute $\ord{\as^3}$ contribution of the leading-power singular
N$^3$LL $\Tau_1$ resummed-expanded cumulant
calculated in~\refcite{Alioli:2023rxx} as $\df \Sigma_{1,\textrm{sub.}}^{\delta\textrm{NNLO}}/ \df \Phi_1$.
In order to regulate the divergencies present at Born level, we impose the generation cuts
\begin{align}
  \label{eq:gencuts1j}
\thetaPSiso(\Phi_M) =
\theta\big(q_T(\Phi_M) > 10^{-3}~\mathrm{GeV}\big) \,,
\end{align}
with $q_T$ the vector boson transverse momentum. We also  use a dynamical cut
\begin{align}
  \Tau_\delta(\Phi_{M}) &=  {\rm min}\left\{
  \Tau_0(\Phi_{M})/2, f\big(q_T(\Phi_{M})\big) \right\}\,,
\end{align}
such that the kinematical limit $\Tau_1(\Phi_2) < \Tau_0(\Phi_2) /2 $ is automatically enforced and the logistic function $f(x)$, defined as
\begin{align}
  f(x) = 10^{-4} \cdot \left( 1 + 10^{-2 + 4/(1 + \exp( -0.3 x))^5}\right)\,,
\end{align}
interpolates the effective $\Tau_\delta$ cutoff between $10^{-4} \leq \Tau_\delta \leq 10^{-2}$ for $5\GeV \lesssim q_T \lesssim 25\GeV$.
In this way $ \Tau_\delta(\Phi_{M}) $ captures the multi-scales dynamics in terms of a single, but kinematic-dependent, resolution parameter.
We therefore arrive at
\begin{align}
  \label{eq:z1slice}
  \Obs_{\delta\textrm{NNLO}}& (\Phi_1)  =   \left. \frac {
		\df
		\Sigma_{1}^{\textrm{N}^3\textrm{LL}}}{
		\df \Phi_1} \big( \Tau_\delta(\Phi_{1}) \big) \right|_{\as^3} \!\thetaPSiso(\Phi_1)  \,\Obs(\Phi_{1})  \\
	&  +   \int  \frac {\df \Phi_{2}}{\df \Phi_{1}} \bigg\{ \,\thetaPSiso(\Phi_2) \,\theta\big(\Tau_1 (\Phi_{2}) > \Tau_\delta(\Phi_{2}) \big) \bigg [
	R\!V(\Phi_{2})    \Obs(\Phi_{2})
	\nn \\ & \qquad \qquad  +  \int  \frac {\df \Phi_{3}}{\df \Phi_{2}}
	\big[ R\!R\,  \Theta_{2}\, \thetaPSiso \big] (\Phi_{3})   \Obs(\Phi_{3})  \theta\big(\Tau_1 (\Phi_{3}) > \Tau_\delta(\Phi_{3}) \big)
	\bigg ] \bigg \}   \nn \\
  &   + \int  \frac {\df \Phi_{3}}{\df \Phi_{1}} \, \bigg[   \big[R\!R\,
       \Theta_{2}\, \thetaPSiso \big] (\Phi_{3}) \Big( \thetaBarProj (\widetilde \Phi_{2}) +  \theta\big(\Tau_1 (\widetilde \Phi_{2}) < \Tau_\delta (\widetilde \Phi_{2})\big) \Big)    \nn\\
     &  \qquad  \qquad \quad +  \big[R\!R\,
          \Theta_{3}\,  \thetaPSiso \big] (\Phi_{3})  \bigg] \ \Obs(\Phi_{3}) \theta\big(\Tau_1 (\Phi_{3}) >
       \Tau_\delta (\Phi_{3}) \big)\,. \nn
\end{align}
Since we choose to perform the $\Phi_2 \to \Phi_3$
splitting in the third line using an FKS mapping that does not preserve
neither $\Tau_0$ nor $q_T$, in the second but last line of the formula above
we have to include the
$\thetaBarProj (\widetilde \Phi_{2}) = \theta\big(q_T(\widetilde \Phi_2) < 10^{-3}~\mathrm{GeV}\big)$ (\emph{i.e.} the complement of \eq{gencuts1j}) and the $\theta\big(\Tau_1 (\widetilde
\Phi_{2}) < \Tau_\delta (\widetilde \Phi_{2})\big) $ contributions, that
exactly accounts for this mismatch.

\begin{figure*}[t!]
  \centering
  \begin{subfigure}[b]{\rescaletwoplots}
    \includegraphics[width=\textwidth]{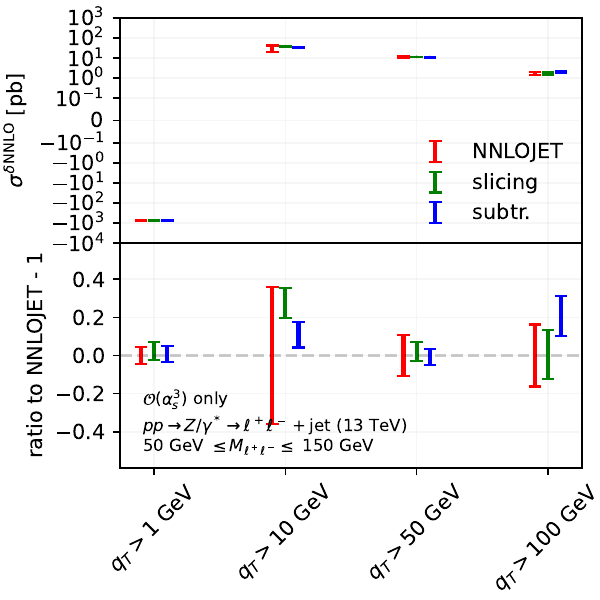}%
  \end{subfigure}
  \begin{subfigure}[b]{\rescaletwoplots}
    \includegraphics[width=\textwidth]{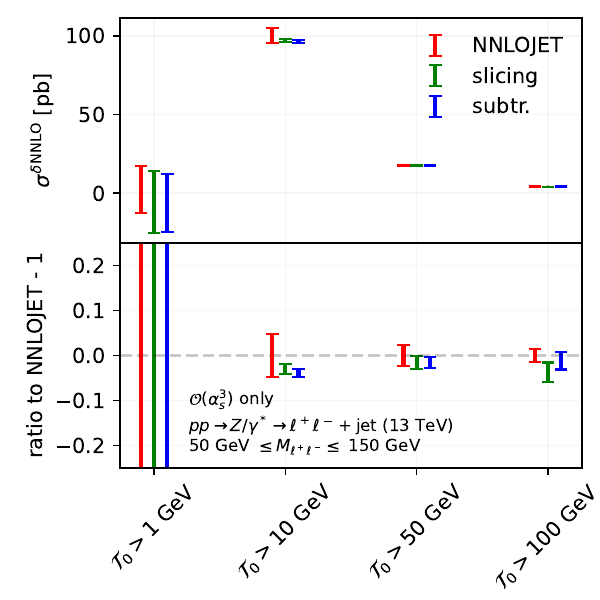}%
  \end{subfigure}
   \caption{Comparison of the $Z$+jet $\ord{\as^3}$ integrated corrections between the slicing and subtraction implementations for  different values of the defining cut on $q_T$ (left) and $\Tau_0$ (right).}
  \label{fig:zjetxs}
\end{figure*}

\begin{figure*}[ht!]
  \centering
  \begin{subfigure}[b]{\rescaletwoplots}
    \includegraphics[width=\textwidth]{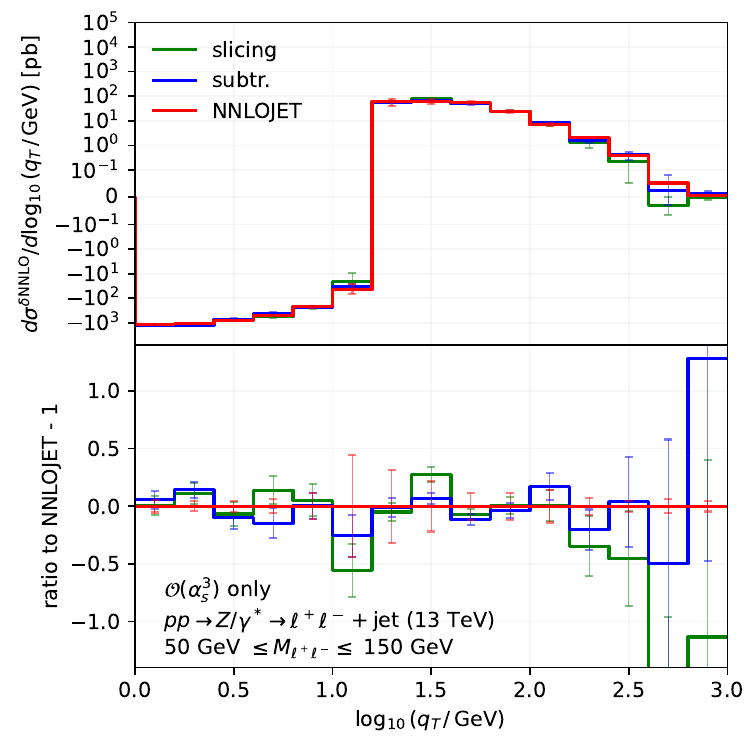}%
  \end{subfigure}
  \begin{subfigure}[b]{\rescaletwoplots}
    \includegraphics[width=\textwidth]{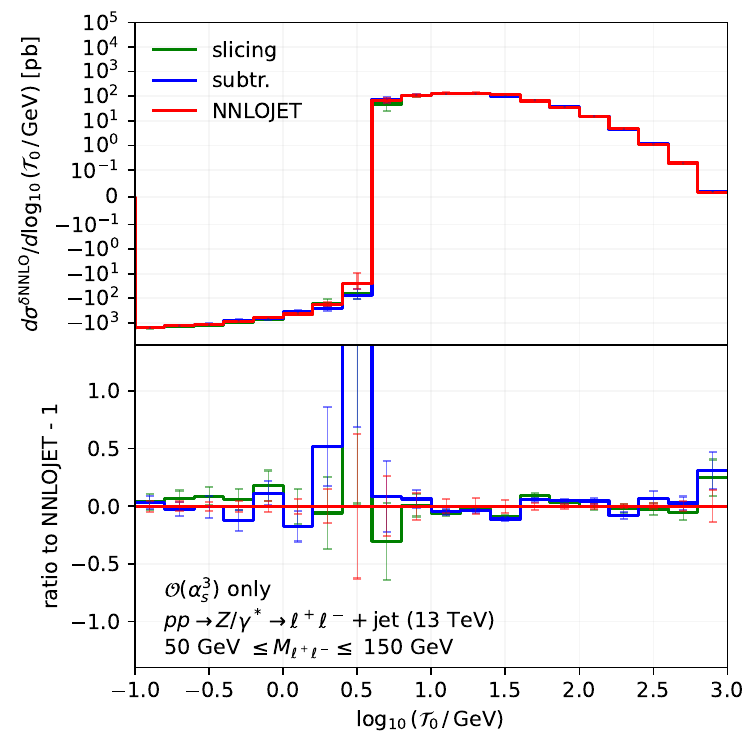}%
  \end{subfigure}
  \caption{Comparison of the $Z$+jet $\ord{\as^3}$ corrections for the $q_T$ (left) and $\Tau_0$ (right) differential distributions between the slicing and subtraction implementations.}
  \label{fig:zjetslicvssub}
\end{figure*}

In the next step we have extended \eq{z1slice} to include a nonlocal
subtraction. The subtraction term can again be retrieved as the leading-power
N$^3$LL $\Tau_1$ resummed-expanded spectrum from~\refcite{Alioli:2023rxx}.
As explained in \sec{gencuts}, the splitting function $P(\Phi_2)$ and the associated mappings are required to preserve
both the $q_T-$defining phase space restriction and the $\Tau_0$ entering the
dynamic cut. We therefore choose to use the \mbox{$(\Tau_0,q_T)$-preserving} mapping
first implemented in \geneva in~\refcite{Alioli:2015toa}. Similarly as for $\Tau_\delta$,  also
$\Tau_1^{\textrm{cut}}$ is now promoted to be a dynamical cut by
\begin{align}
  \Tau_1^{\textrm{cut}}(\Phi_M) = {\rm min}\left\{\Tau_0(\Phi_{M})/2, 10^2 \cdot f\big(q_T(\Phi_{M})\big) \right\}.
  \end{align}
With these choices, we get the formula for the NNLO corrections with a nonlocal subtraction  as
\begin{align}
  \label{eq:z1subtr}
    \Obs_{\delta\textrm{NNLO}}& (\Phi_1)  =   \left. \frac {
		\df
		\Sigma_{1}^{\textrm{N}^3\textrm{LL}}}{
		\df \Phi_1} \big( {\Tau_1^{\textrm{cut}}(\Phi_{1})} \big) \right|_{\as^3} \!\thetaPSiso(\Phi_1)  \,\Obs(\Phi_{1})  \\
	&  +   \int  \frac {\df \Phi_{2}}{\df \Phi_{1}} \bigg\{ \thetaPSiso(\Phi_{2}) \theta\big(\Tau_1 (\Phi_{2}) > \Tau_\delta(\Phi_{2}) \big) \bigg [
	R\!V(\Phi_{2})    \Obs(\Phi_{2})
	\nn \\ & \qquad \qquad  +  \int  \frac {\df \Phi_{3}}{\df \Phi_{2}}
	\big[ R\!R\, \Theta_{2} \,\thetaPSiso \big] (\Phi_{3})   \Obs(\Phi_{3})  \theta\big(\Tau_1 (\Phi_{3}) > \Tau_\delta(\Phi_{3}) \big)   \nn \\
         	& \ \  - \left. \frac {\df \sigma_{1}^{\textrm{N}^3\textrm{LL}}}{\df
		\Phi_1 \df \Tau_1} \right|_{\as^3} \ P(\Phi_{2}) \Obs(\Phi_{1}) \theta\big(  \Tau_1 (\Phi_{2}) <
	{\Tau_1^{\textrm{cut}}(\Phi_{2})}\big)   \bigg ] \bigg \}   \nn \\
  &   + \int  \frac {\df \Phi_{3}}{\df \Phi_{1}} \bigg[   \big[R\!R\,
       \Theta_{2} \,\thetaPSiso \big] (\Phi_{3}) \Big( \thetaBarProj (\widetilde \Phi_{2}) +  \theta\big(\Tau_1 (\widetilde \Phi_{2}) < \Tau_\delta (\widetilde \Phi_{2})\big) \Big)    \nn\\
     &  \qquad  \qquad \quad +  \big[R\!R\,
          \Theta_{3} \,\thetaPSiso \big] (\Phi_{3})  \bigg] \ \Obs(\Phi_{3}) \theta\big(\Tau_1 (\Phi_{3}) >
       \Tau_\delta (\Phi_{3}) \big)\,. \nn
\end{align}
It must be pointed out, however,  that in the region where $\Tau_0(\Phi_{M}) \leq 2 f\big(q_T(\Phi_{M})\big)$ the
nonlocal subtraction calculation becomes a pure slicing one, because $\Tau_1^{\textrm{cut}} = \Tau_\delta = \Tau_0(\Phi_{M})/2$.

In \Fig{zjetxs} we compare the results of the slicing and subtraction approaches described above with \nnlojet
for the $\ord{\as^3}$ contributions to the cross sections with different defining cuts in $q_T$ and $\Tau_0$. We observe good agreement for all values of the cuts, with the results of the subtraction approach giving statistical errors consistently smaller than the corresponding ones obtained with the slicing approach.

In a similar fashion,  we compare the $q_T$ and $\Tau_0$ distributions in \Fig{zjetslicvssub}. We observe a similar good agreement  with \nnlojet over several orders of magnitude, for both the slicing and subtraction approaches. The statistical errors are now more challenging, for both \geneva and \nnlojet, especially where the distribution changes sign.

%%%%%%%%%%%%%%%%%%%%%%%%%%%%%%%%%%%%%%%%%%%%%%%%%%%%%%%%%%%%%%%%%%%%%%%%%%%%%%%%
\subsubsection{Inclusion of FPC corrections}
\label{sec:zj_fpc}
%%%%%%%%%%%%%%%%%%%%%%%%%%%%%%%%%%%%%%%%%%%%%%%%%%%%%%%%%%%%%%%%%%%%%%%%%%%%%%%
In order to include the fiducial power corrections below the IR cutoff, we can specify the general \eq{finalP2B} to the one-jettiness case, obtaining
\begin{align}
  \label{eq:finalP2Bzj}
	\Obs_{\delta\textrm{NNLO}}& (\Phi_1)  =  \Bigg\{  \left. \frac {
		\df
		\Sigma_{1}^{\textrm{N}^3\textrm{LL}}}{
		\df \Phi_1} \big( \Tau_1^{\textrm{cut}}(\Phi_1)\big) \right|_{\as^3} \! \thetaPSiso(\Phi_1)   \\
	&  +   \int  \frac {\df \Phi_{2}}{\df \Phi_1} \bigg\{ \thetaPSiso(\Phi_{2})  \theta\big(\Tau_1 (\Phi_{2}) > \Tau_\delta(\Phi_{2}) \big) \bigg [
	R\!V(\Phi_{2})
	\nn \\ & \qquad \qquad   +  \int  \frac {\df \Phi_{3}}{\df \Phi_{2}}
	\big[ R\!R\  \Theta_{2} \thetaPSiso \big] (\Phi_{3})   \theta\big(\Tau_1 (\Phi_{3}) > \Tau_\delta(\Phi_{3}) \big)  \nn\\
	& \qquad \qquad  - \left. \frac {\df \sigma_{1}^{\textrm{N}^3\textrm{LL}}}{\df
		\Phi_1 \df \Tau_1} \right|_{\as^3} \ P(\Phi_{2})  \theta\big(  \Tau_1 (\Phi_{2}) <
	{\Tau_1^{\textrm{cut}}(\Phi_{2})}\big) \bigg ]    \bigg \}   \nn \\
  &   + \int  \frac {\df \Phi_{3}}{\df \Phi_1} \Bigg[   \big[R\!R\
       \Theta_{2} \thetaPSiso \big] (\Phi_{3}) \Big( \thetaBarProj (\widetilde \Phi_{2}) +  \theta\big(\Tau_1 (\widetilde \Phi_{2}) < \Tau_\delta (\widetilde \Phi_{2})\big) \Big)    \nn\\
     &  \qquad  \qquad \quad +  \big[R\!R\
          \Theta_{3} \thetaPSiso \big] (\Phi_{3})  \Bigg]  \theta\big(\Tau_1 (\Phi_{3}) >
       \Tau_\delta (\Phi_{3}) \big)  \Bigg \}\ \Obs(\Phi_1)\  \nn\\
 &  +   \int  \frac {\df \Phi_{2}}{\df \Phi_1} \thetaPSiso(\Phi_{2}) \bigg [
          R\!V(\Phi_{2}) \big[  \Obs(\Phi_{2}) { - \Obs(\Phi_1)}\big]
          \nn \\ & \qquad \qquad   + \int  \frac {\df \Phi_{3}}{\df \Phi_{2}}
          \big[ R\!R\ \Theta_{2}  \thetaPSiso \big] (\Phi_{3})  \big[
                   \Obs(\Phi_{3}) { -
                   \Obs(\Phi_1)}\big]  \bigg] \nn\\
     &  +   \int  \frac {\df \Phi_{3}}{\df \Phi_1} \thetaPSiso(\Phi_{3}) \bigg[   \big[R\!R\
       \Theta_{2} \big] (\Phi_{3}) \thetaBarProj (\widetilde
       \Phi_{2}) + \big[R\!R\
          \Theta_{3} \big] (\Phi_{3}) \bigg] \nn\\ & \qquad \qquad \qquad \qquad
                                                         \times \ \big[
       \Obs(\Phi_{3}) { -  \Obs(\Phi_1)}\big]  \nn\,.
\end{align}
As discussed after \eq{finalP2B}, the formula above is only correct
assuming the $\Phi_3 \to \Phi_2 \to \Phi_1 $ projections, which are
needed for the evaluation of the subtracted observable according to
the P2B method, are performed with a mapping that preserves the phase
space restrictions at the generation level, \emph{i.e.}
$\thetaPSiso(\Phi_{3}) = \thetaPSiso(\Phi_{2}) = \thetaPSiso(\Phi_{1})
= \theta\big(q_T(\Phi_M) > 10^{-3}~\mathrm{GeV}\big)$. This is a very
nontrivial requirement that forced us to design a new $q_T-$preserving
mapping, which is applied recursively.

\begin{paragraph}{P2B mapping}
We first find the closest
partons comparing an invariant mass metric for final-state distances
and the transverse momentum squared with respect to the beam for the
initial-state distances, dubbing them emitted and sister.
In the case of a final-state clustering,
the spatial three-vector momenta of the closest partons are summed together
and the energy of the clustered parton is set equal to the modulus of
this sum, thus replacing two massless partons with a new massless one,
which is dubbed the mother.
All the other final-state partons and colour-neutral objects are left
unchanged. In formulae
\begin{align}
  \tilde p_i = p_i \quad \textrm{for} \  i \neq \textrm{emi.}\,, \textrm{sis.} \qquad \textrm{and} \qquad
  \tilde p_{\textrm{mum}} = \left( | \vec{p}_{\textrm{emi.}} + \vec{p}_{\textrm{sis.}}| , \vec{p}_{\textrm{emi.}} + \vec{p}_{\textrm{sis.}} \right) \nn\,.
\end{align}
This implies that the total energy and rapidity of the event are modified because, calling
the sum of the four-momenta in the final state before and after the clustering as
\begin{align}
  p_\textrm{fin.} = \sum_{i \in \textrm{fin. state}} p_i \quad \textrm{and} \quad  \tilde p_\textrm{fin.} = \sum_{i \in \textrm{fin. state}} \tilde p_i \,,\quad \textrm{respectively,} \nn
  \end{align}  we
observe that $\tilde s = \tilde p_{\textrm {fin.}}^2 \neq p_{\textrm {fin.}}^2  $ and
$ \tilde Y = \frac{1}{2} \log\left(\frac{\tilde p_{\textrm {fin.}}^0 + \tilde p_{\textrm {fin.}}^z}{\tilde p_{\textrm {fin.}}^0 - \tilde p_{\textrm {fin.}}^z}\right) \neq Y$.
This  is accounted for by rescaling the colliding partons' energies and momentum fractions
as follows
\begin{align}
  \label{eq:p2bres}
  \tilde p_a = \left(\frac{1}{2} \sqrt{\tilde s} e^{\tilde Y}, 0, 0, \frac{1}{2} \sqrt{\tilde s} e^{\tilde Y}\right)\,,  \qquad
  \tilde p_b = \left(\frac{1}{2} \sqrt{\tilde s} e^{\tilde Y}, 0, 0, -\frac{1}{2} \sqrt{\tilde s} e^{\tilde Y}\right).
  \end{align}
In the case of an initial-state clustering instead, one removes the final state parton that is closer to the beam (\emph{i.e.} the one with smallest transverse momentum, dubbed emitted).
The colour-neutral particles in the final state and all the other coloured partons but the one that
has the smallest transverse momentum among the remaining ones, called next-to-softest (nts.), are instead left unchanged.
It is in fact necessary to modify the transverse momentum of at least one of the remaining partons in order to absorb the transverse momentum
of the removed emission, without modifying the colour-neutral system.
The $z$-component of this parton is preserved and  its energy is corrected
to maintain it on-shell. In formulae
\begin{align}
  \tilde p_i = p_i \quad \textrm{for} \  i \neq \textrm{emi.}\,, \textrm{nts.}\qquad \textrm{and} \qquad
  \tilde p_{\textrm{nts.}} = \bigg( \tilde p_\textrm{nts.}^0 , - \sum_{i \neq \textrm{nts.}} \tilde p_i^x , - \sum_{i \neq \textrm{nts.}} \tilde p_i^y   , p_{\textrm{nts.}}^z\bigg) \nn\,,
\end{align}
where $\tilde p_\textrm{nts.}^0  = \sqrt{ (\tilde p_\textrm{nts.}^x)^2 + (\tilde p_\textrm{nts.}^y)^2 +  (p_{\textrm{nts.}}^z)^2}$.
At this point one can again rescale the initial state momenta as in \eq{p2bres} to accommodate for the change in
the total energy and rapidity of $\tilde p_{\textrm {fin.}}$.
We point out that both these mappings actually preserve the full four-momentum of the colour-singlet system,
not only its transverse components. This means that any observable which is entirely determined by the momenta
of the colour-singlet system or by its decay product (\emph{e.g.} the leptons coming from the $Z$ decay) will exactly be
preserved by the projection and will therefore not receive any additional fiducial power correction.
\end{paragraph}
\\~\\

\begin{figure*}[t!]
  \centering
  \begin{subfigure}[b]{\rescaletwoplots}
    \includegraphics[width=\textwidth]{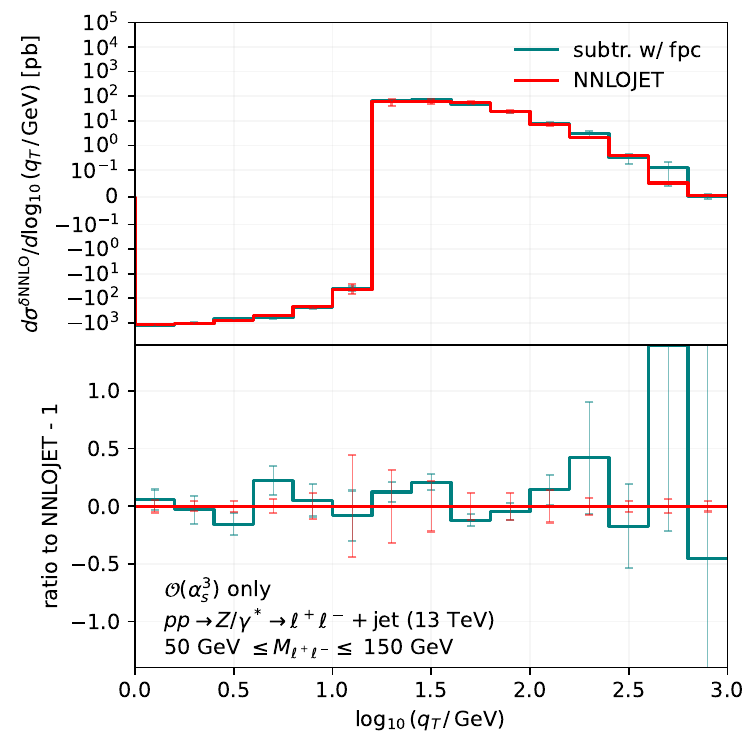}%
  \end{subfigure}
  \begin{subfigure}[b]{\rescaletwoplots}
    \includegraphics[width=\textwidth]{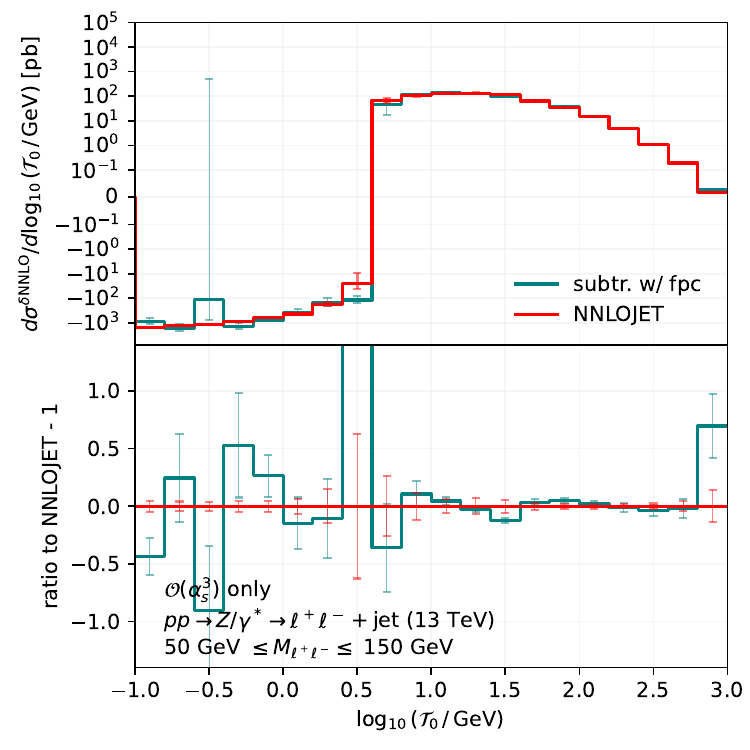}%
  \end{subfigure}
  \caption{Comparison with \nnlojet of the $Z$+jet $\ord{\as^3}$ corrections for the $q_T$ (left) and $\Tau_0$ (right) differential distributions within the subtraction implementation including FPCs.}
\label{fig:zjetslicvssubfpc}
\end{figure*}

\begin{figure*}[ht!]
  \centering
  \begin{subfigure}[b]{\rescaletwoplots}
    \includegraphics[width=\textwidth]{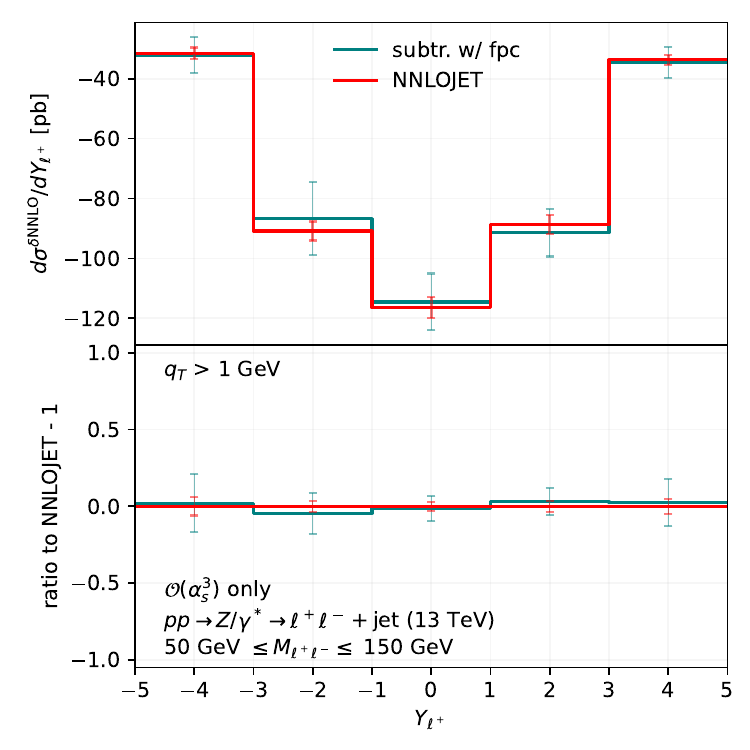}%
  \end{subfigure}
  \begin{subfigure}[b]{\rescaletwoplots}
    \includegraphics[width=\textwidth]{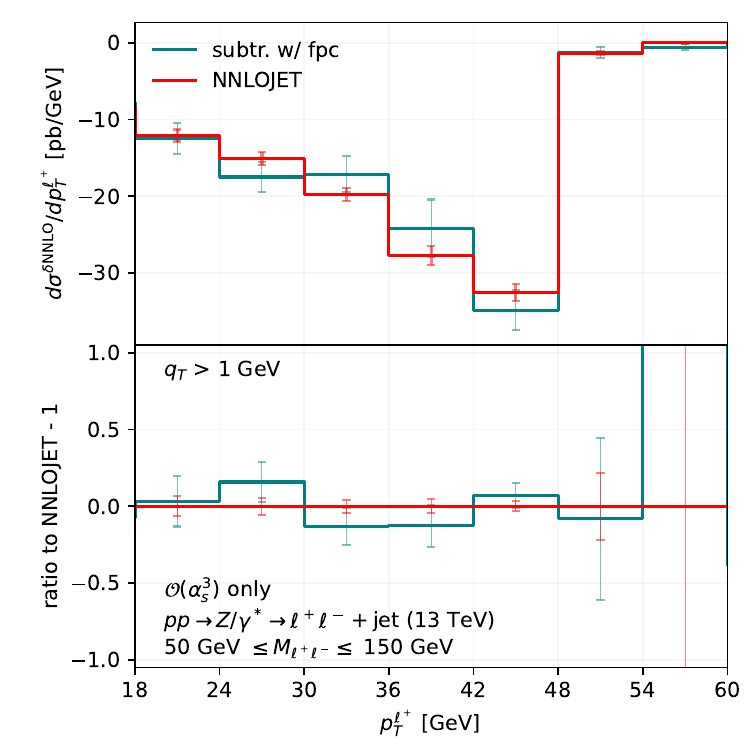}%
  \end{subfigure}
  \caption{Comparison with \nnlojet of the $Z$+jet $\ord{\as^3}$ corrections for the rapidity $Y_{\ell^+}$ (left) and transverse momentum $p_T^{\ell^+}$ (right) differential distributions  of the positively charged lepton with a cut $q_T > 1$~GeV,  within the subtraction implementation including FPCs.}
\label{fig:zjetlep1}
\end{figure*}

\begin{figure*}[ht!]
  \centering
  \begin{subfigure}[b]{\rescaletwoplots}
    \includegraphics[width=\textwidth]{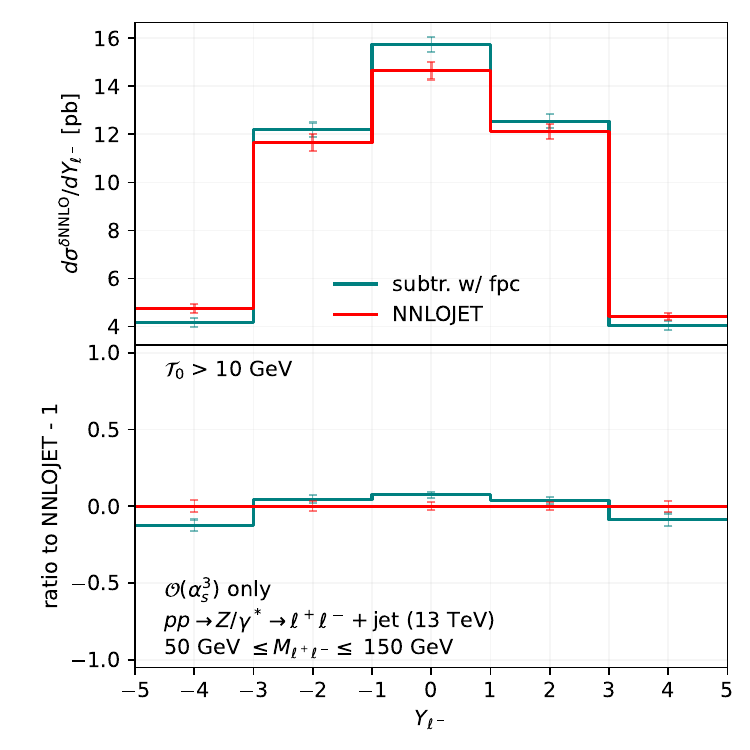}%
  \end{subfigure}
  \begin{subfigure}[b]{\rescaletwoplots}
    \includegraphics[width=\textwidth]{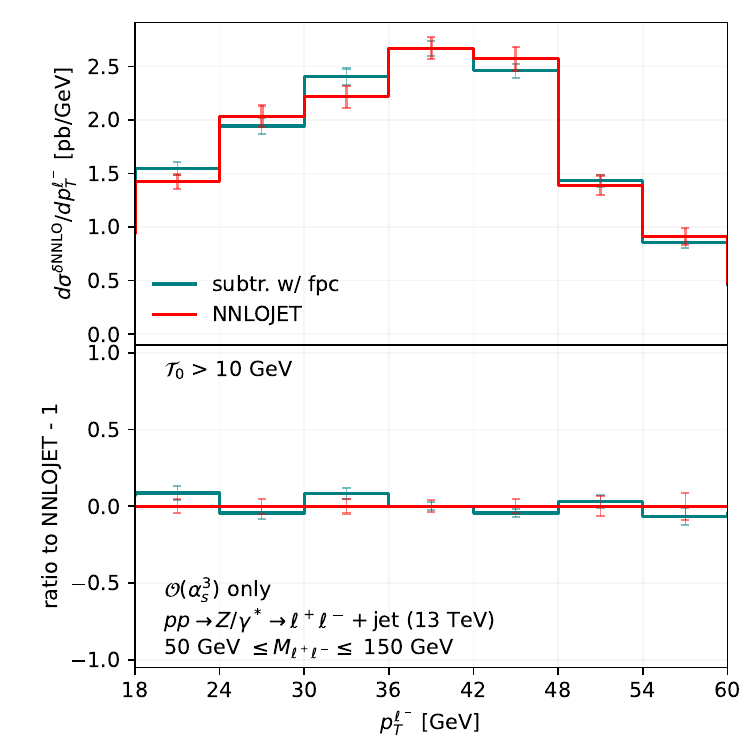}%
  \end{subfigure}
  \caption{Comparison with \nnlojet of the $Z$+jet $\ord{\as^3}$ corrections for the rapidity $Y_{\ell^-}$ (left) and transverse momentum $p_T^{\ell^-}$ (right) differential distributions  of the negatively charged lepton with a cut $\Tau_0 > 10$~GeV,  within the subtraction implementation including FPCs.}
\label{fig:zjetlep2}
\end{figure*}

In \Fig{zjetslicvssubfpc} we compare the $\ord{\as^3}$ corrections obtained by \nnlojet to those obtained by \geneva with the nonlocal subtraction and including the FPCs    for the $q_T$  and $\Tau_0$ distributions.  We observe the same good level of agreement both for an observable like $q_T$ which is preserved by the P2B mapping as well as for $\Tau_0$, which is instead not preserved and subject to FPCs.
In a similar fashion we compare the distributions of the decay products of the vector boson with a cut in $q_T$ in \Fig{zjetlep1} and with a cut in $\Tau_0$ in \Fig{zjetlep2}. Due to the complete preservation of the entire four momentum of the vector boson and of its decay products in the P2B mapping, the complete distributions in \Fig{zjetlep1} are preserved. This is not the case for the distributions in \Fig{zjetlep2}, due to the cut on $\Tau_0$. We observe nonetheless a very good agreement with \nnlojet for both the cases.

We also notice a difference in the size of the statistical errors on the leptonic distributions compared to the one on the inclusive cross section in \Fig{zjetxs} for the same value of the cut in $q_T$ or $\Tau_0$.
Indeed, by calculating the \nnlojet cross section integrating over the bins of the leptonic distributions, one would find compatible results with a smaller error.
This is likely a consequence of the \nnlojet treatment of outliers, as described in~\cite{NNLOJET:2025rno,Gehrmann-DeRidder:2016ycd}, which is applied on a bin-to-bin basis. On the other hand, \geneva does apply a trimming procedure to discard runs classified as outliers, but at the level of the fiducial cross section only.
This preserves the statistical error of the integrated cross section, at the expense of larger errors in leptonic distributions.

%%%%%%%%%%%%%%%%%%%%%%%%%%%%%%%%%%%%%%%%%%%%%%%%%%%%%%%%%%%%%%%%%%%%%%%%%%%%%%%%
\section{Extension to higher orders}
\label{sec:n3lo}
\subsection{Validation of $q_T$ and $\Tau_0$ spectra against $\ord{\as^3}$ singular predictions}
\label{sec:zjet_fo_vs_resexp}

A crucial benefit of nonlocal subtractions is that the QCD IR singularities, present in
the intermediate steps of a calculation, are mapped to a single variable and the counterterms are given
in terms of the singular limits of physical cross sections.
This renders such methods particularly appealing for higher-order predictions, because they are only limited
by the availability of the perturbative ingredients that enter the `subtraction' cross section.
This is opposed to local subtractions, where disentangling the various (overlapping) singular regions and constructing appropriate (integrated) subtraction
counter-terms has been proved to be a formidable task, although recent improvements suggest that their full automation at NNLO might not be far away~\cite{Fox:2024bfp, Devoto:2025kin,Bertolotti:2022aih}.

\begin{figure*}[t!]
  \centering
  \begin{subfigure}[b]{\rescaletwoplots}
    \includegraphics[width=\textwidth]{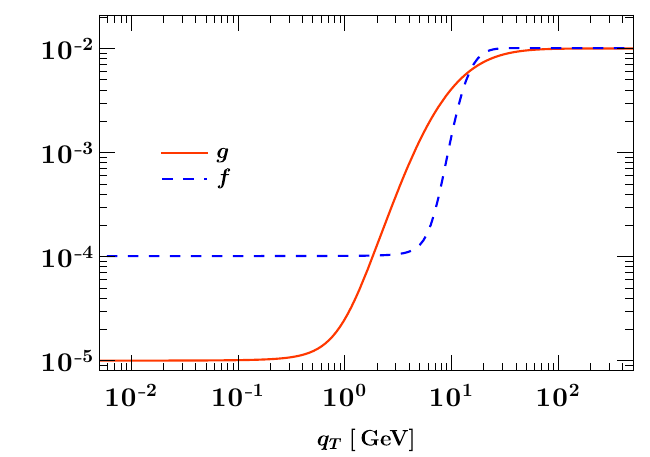}%
  \end{subfigure}
  \begin{subfigure}[b]{\rescaletwoplots}
    \includegraphics[width=\textwidth]{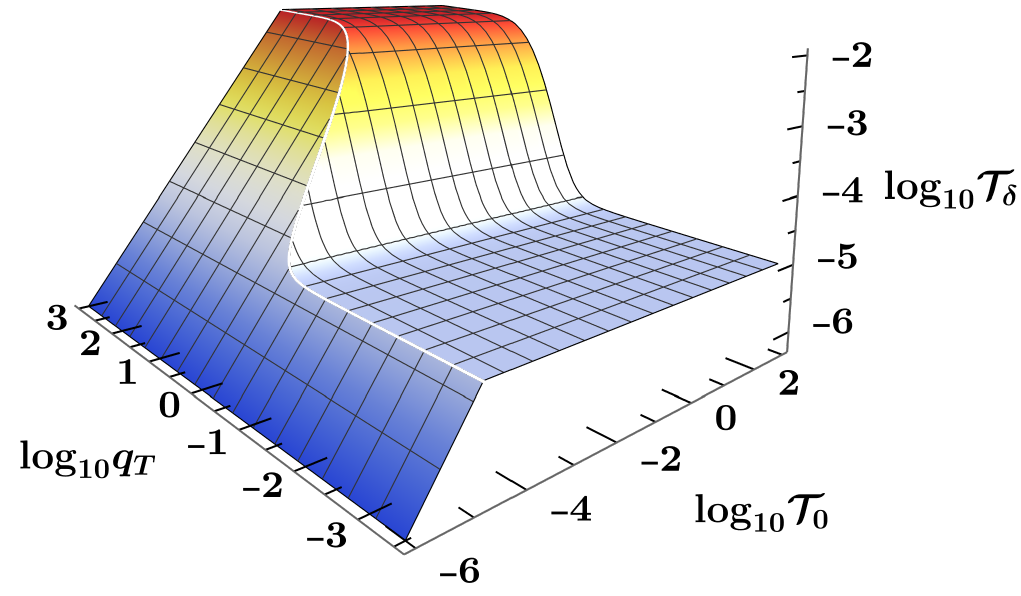}%
  \end{subfigure}
  \caption{Comparison of the functional forms for the dynamical cut (left) and  complete dependence of the new $\Tau_\delta$ on $\Tau_0$ and $q_T$ (right).}
\label{fig:tau_delta}
\end{figure*}

\begin{figure*}[ht!]
  \centering
  \begin{subfigure}[b]{\rescaletwoplots}
    \includegraphics[width=\textwidth]{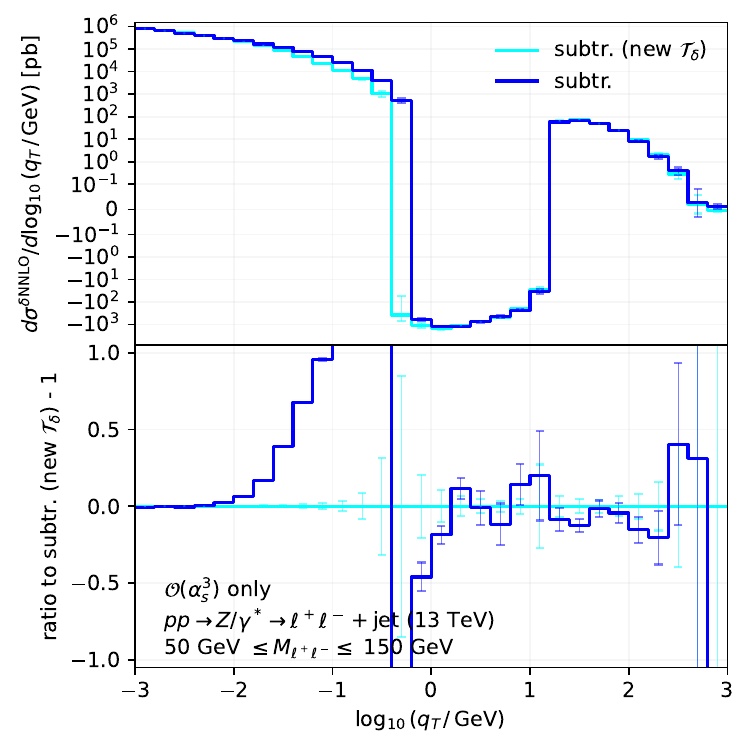}%
  \end{subfigure}
  \begin{subfigure}[b]{\rescaletwoplots}
    \includegraphics[width=\textwidth]{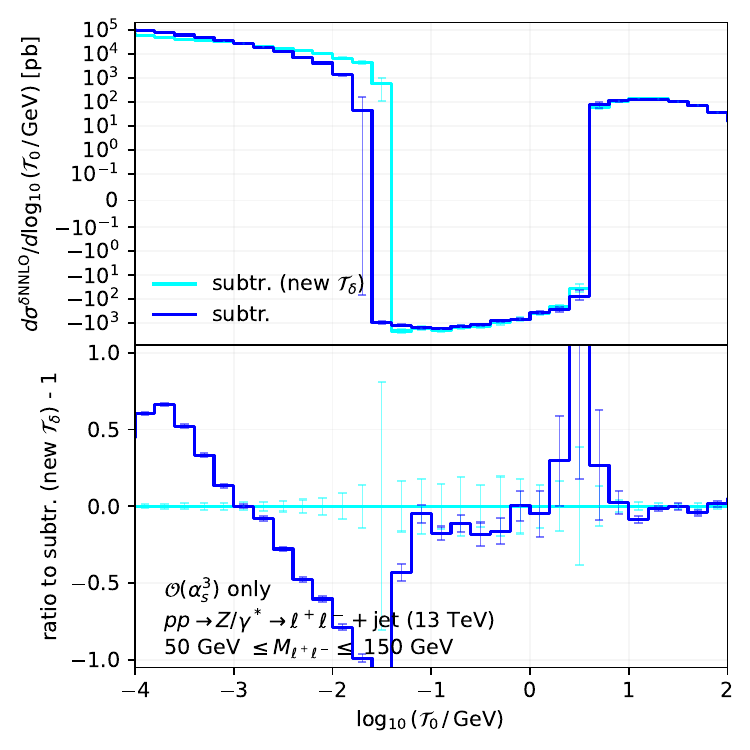}%
  \end{subfigure}
  \caption{Comparison of the \geneva results for the $Z$+jet $\ord{\as^3}$ corrections for the $q_T$ (left) and $\Tau_0$ (right) differential distributions between different choices for the dynamical infrared cutoff within a subtraction implementation.}
\label{fig:cmp_tau_delta}
\end{figure*}

On the other hand, it is a known fact that the residual dependence on the technical cut-off $\Tau_\delta$ in nonlocal subtractions affects the lower range of validity for the predictions of observables that are considered to be hard scales of the process.
Two examples are the transverse momentum spectrum ($q_T$) of the colour-singlet system
and the zero-jettiness ($\Tau_0$) resolution variables that, when they are evaluated at extremely small values, show a strong dependence on $\Tau_{\delta}$. Thus, as part of our NNLO validation for the $Z$+jet process,
it is of outmost importance to quantitatively study the limitations of the \geneva predictions when employing a certain cut-off value $\Tau_\delta$
in the $\Tau_1$ subtractions.
In particular, in order to extend the value of $\Tau_\delta$ to lower values, while avoiding to violate the assumptions behind the factorization theorem for $\Tau_1$, which requires $\Tau_1$ to be much smaller than any other scale in the problem, including $\Tau_0$ or $q_T$, here we explore the possibility to change the definition
\begin{align}
  \Tau_\delta(\Phi_{M}) &=  {\rm min}\left\{
  \Tau_0(\Phi_{M})/2, g\big(q_T(\Phi_{M})\big) \right\} \  \textrm{with}\ g(x) &= 10^{-5} \cdot \left( 1 + 10^{-3 + 6/(1 + \exp( -2 \sqrt[3]{x}))^5}\right)\,,\nn
  \end{align} which  interpolates the effective $\Tau_\delta$  between $10^{-5}$ and $10^{-2}$ when $q_T$ varies between $0$ and $100$~GeV. In the left panel of \Fig{tau_delta} we show the comparison between the functional forms for the dynamical cut on $\Tau_\delta$ used in the previous section, compared to the new form, which can reach the smaller value of $10^{-5}$~GeV for values of $q_T \lesssim 1 $~GeV. In the right panel of the same figure we show the new full $\Tau_\delta$ dependence on both $\Tau_0$ and $q_T$. As for the previous functional form, when $\Tau_0(\Phi_M) < 2 g (q_T(\Phi_M))$  one has $\Tau_\delta=\Tau_1^\cut = \Tau_0 /2$ and the nonlocal subtraction falls back to be a pure slicing calculation.
We compare the results for this new choice of the infrared cutoff with those used for the predictions in the previous sections in \Fig{cmp_tau_delta}.

\begin{figure*}[ht!]
  \centering
  \begin{subfigure}[b]{\rescaletwoplots}
    \includegraphics[width=\textwidth]{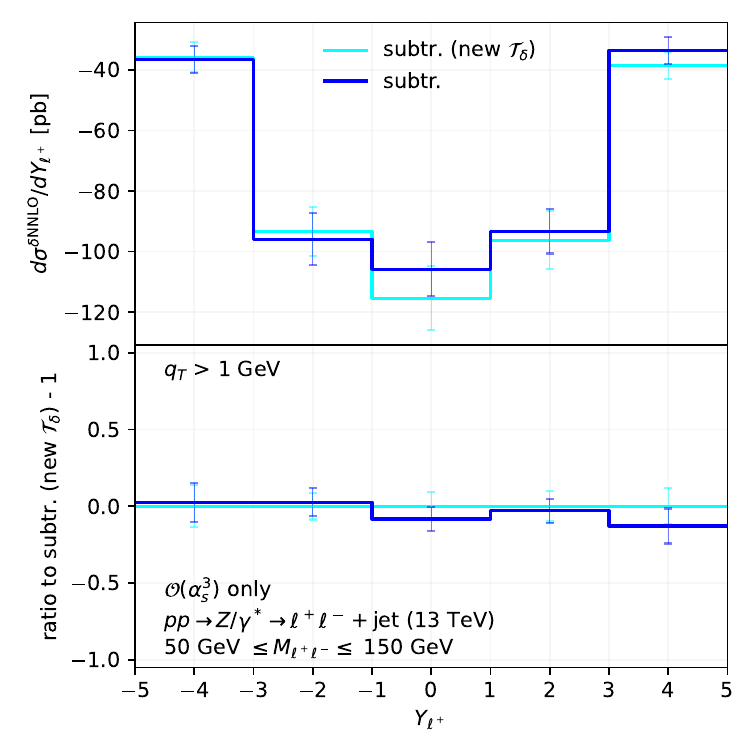}%
  \end{subfigure}
  \begin{subfigure}[b]{\rescaletwoplots}
    \includegraphics[width=\textwidth]{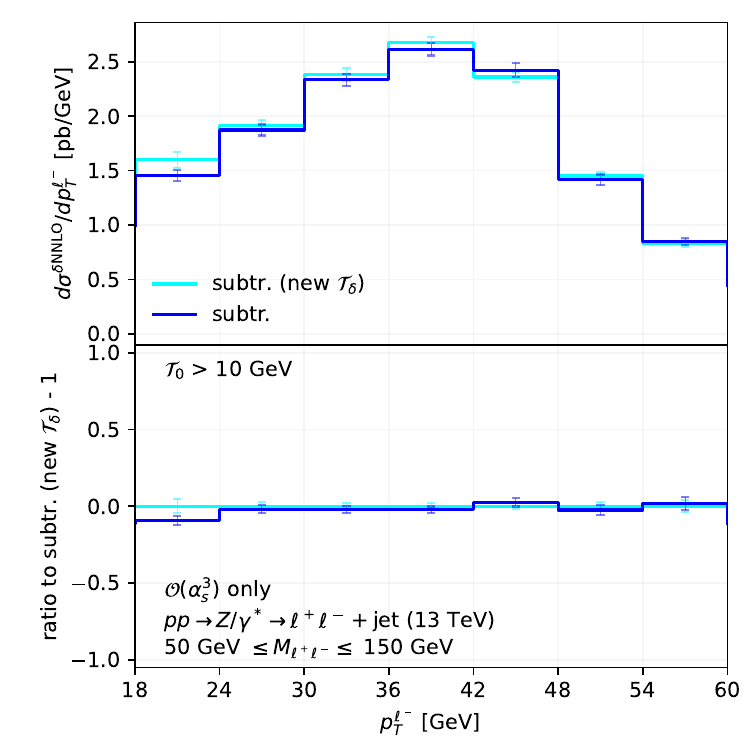}%
  \end{subfigure}
  \caption{Comparison of the $Z$+jet $\ord{\as^3}$ differential corrections for the rapidity $Y_{\ell^+}$ (left)  of the positively charged lepton with a cut $q_T > 1$~GeV  and of the transverse momentum $p_T^{\ell^-}$ (right)  of the negatively charged lepton with a cut $\Tau_0 > 10$~GeV,  within the subtraction implementation for different values of the infrared cutoff $\Tau_\delta$.}
\label{fig:zjetleplow}
\end{figure*}

We first notice how the curves are identical for values of $q_T \gtrsim 1$~GeV and $\Tau_0 \gtrsim 0.1$~GeV, thus ensuring that the dynamical cutoff employed in the previous sections is correct for the observables studied there.
This is further confirmed  by the comparison of the differential distribution of the $Z$ decay products in \Fig{zjetleplow}.

For smaller values of $q_T$ or $\Tau_0$, however, we notice a  considerable shift between the predictions obtained with the different dynamical cuts.  This behaviour is expected, because for those small values of $q_T$ or $\Tau_0$ the previous form of the infrared cutoff $\Tau_\delta$  enters a kinematic regime where the factorization theorem for $\Tau_1$ is no longer valid.

\begin{figure*}[ht!]
  \centering
  \begin{subfigure}[b]{\rescaletwoplots}
    \includegraphics[width=\textwidth]{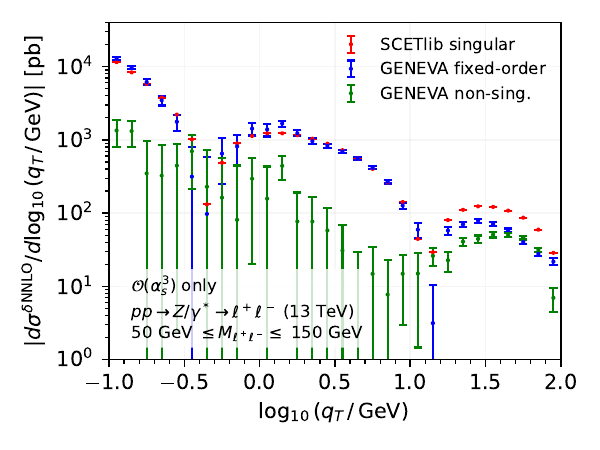}%
  \end{subfigure}
  \begin{subfigure}[b]{\rescaletwoplots}
    \includegraphics[width=\textwidth]{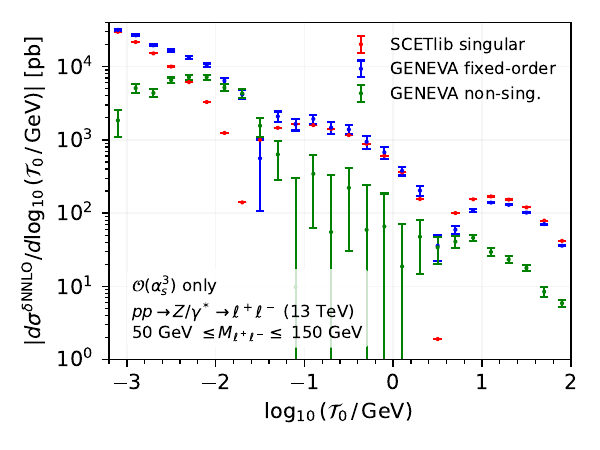}%
  \end{subfigure}
  \caption{Validation of \geneva $\ord{\as^3} \,\, Z+$jet predictions in small-$q_T$ (left) and small-$\Tau_0$ (right) regions against singular predictions from \scetlib
  with the nonsingular contribution (green) corresponding to their difference.}
\label{fig:zjet_n3lo_coeff_nons}
\end{figure*}

In this situation one might ask how to determine the correct form for the dynamical cut and how low one can push the infrared cutoff, considering that when the cutoff is pushed to extremely low values the NLO calculation above the cut becomes very unstable, being evaluated close to its IR limits.
The correct way to address this question is to use as IR cutoff the largest value that keeps the NLO calculation above the cut stable while capturing the correct singular behaviour of the $q_T$ or $\Tau_0$ spectra.
In order to further explain this point, in \Fig{zjet_n3lo_coeff_nons} we show the $\ord{\as^3}$ coefficient for the $q_T$ (left) and the $\Tau_0$ (right) spectrum
from \geneva (blue) and from \scetlib (red)~\cite{scetlib, Lustermans:2019plv, Billis:2019vxg, Billis:2021ecs, Billis:2024dqq}, where the latter predicts the singular cross section
of the corresponding resolution variable. The singular cross section is known to dominate in the limits $q_T, \Tau_0 \to 0$ due to large,
unresummed Sudakov double-logarithms and its difference to the fixed-order cross section, here denoted as `nonsingular' (green), is expected to show a power suppressed behaviour~\cite{Collins:1981uk, Collins:1981va, Collins:1984kg, Becher:2010tm, Catani:2000vq, GarciaEchevarria:2011rb, Chiu:2012ir, Collins:2012uy, Li:2016axz, Stewart:2009yx, Stewart:2010tn}.

It is evident for both resolution variables that the nonsingular cross section is significantly suppressed for low values of
$q_T$ and $\Tau_0$, although a clear power law is hard to discern. This is partially attributed to the fact that
the coefficients have two zero-crossings deep in the singular region ($q_T \simeq 0.4 \GeV, 10\GeV$ and $\Tau_0\simeq 2\times 10^{-2}\GeV, 3\GeV$),
and it is thus numerically challenging to obtain the expected quadratic (linear) power suppression for $q_T$ ($\Tau_0$) in the region in-between.
In contrast, a downward trend for $\Tau_0 \lesssim 10^{-2}\GeV$ can be seen but the effect of the technical cut-offs of the NLO$_1$ calculation become significant
below $\Tau_0 \lesssim 7 \times 10^{-4} \GeV$, which effectively defines the lowest valid value of the $\Tau_0$ fixed-order results of \geneva for this choice of $\Tau_\delta$ and the NLO technical cuts.
Similar conclusions are also drawn for $q_T$, with the lowest valid value at $q_T \simeq 0.2\GeV$.

We stress though, that employing even lower values of the
technical cut-off $\Tau_\delta$ would allow us in principle to push the range of validity for the $q_T$ and $\Tau_0$ spectra at even smaller values,
but this has to be balanced with the instability of the NLO calculation with reduced technical cuts.
We remind the reader that even for local subtraction methods is computationally challenging to properly describe the small $q_T$ or $\Tau_0$ regions.

Finally, in addition to the zero-crossings of the coefficients,
we remind the known fact that for this process, the aggregate of all partonic channels is known to exhibit significant numerical cancellations
which potentially further obscure the expected power-suppressed trend of the nonsingular. A more careful treatment could involve the inspection of the
nonsingular cross section for each partonic channel separately, as done \emph{e.g.} in \refscite{Moult:2016fqy, Moult:2017jsg, Ebert:2018gsn}.
This will potentially allow us to unravel the channel-specific scaling trend of the nonsingular cross section. We leave this study to future work.

%%%%%%%%%%%%%%%%%%%%%%%%%%%%%%%%%%%%%%%%%%%%%%%%%%%%%%%%%%%%%%%%%%%%%%%%%%%%%%%
\subsection{Extension to N$^3$LO colour-singlet hadroproduction}
Having validated the nonsingular spectrum  against the N$^3$LL singular
predictions and given the availability of all the three-loop boundary terms
necessary for the N$^3$LL$^\prime$ resummation for both $q_T$ and $\Tau_0$ for
colour-singlet hadroproduction~\cite{Baranowski:2024vxg, Ebert:2020unb, Bruser:2018rad}, we can extend  \eq{finalNNLO} to one order
higher, in order to calculate the N$^3$LO fully-exclusive
corrections.
Using $\Tau_0$ as example resolution variable (but the formula applies equally
to $q_T$) we obtain
\begingroup
  \allowdisplaybreaks
\begin{align}
  \label{eq:finalN3LO}
	\Obs_{\delta\textrm{N$^3$LO}} (\Phi_0)  = &  \frac {\df \Sigma_{0,\textrm{sub.}}^{\delta\textrm{N$^3$LO}}}{\df \Phi_0} \big( \Tau_0^{\textrm{cut}}\big)   \,\Obs(\Phi_{0})  \\
	&  + \int  \frac {\df \Phi_{1}}{\df \Phi_{0}} \bigg [ { \frac{ \df \sigma_{1}^{\delta\textrm{NNLO}}}{\df \Phi_{1}}}  - \frac {\df \sigma_{0,\textrm{sub.}}^{\delta\textrm{N$^3$LO}}}{\df \Phi_0\,\df \Tau_0} \ P(\Phi_{1}) \, \theta\big( \Tau_0 (\Phi_{1}) < \Tau_0^{\textrm{cut}}\big)  \bigg]  \nn \\
	& \qquad \qquad \qquad \times  \Obs(\Phi_{0})  \ \theta \Big( \Tau_0(\Phi_{X=\{1,2,3\}}) > \Tau_{0,\delta} \Big)\nn \\
	& + \int \frac {\df \Phi_{1}}{\df \Phi_{0}} \frac{ \df
          \sigma_{1}^{\delta\textrm{NNLO}}}{\df \Phi_{1}} \bigg[
          \Obs(\Phi_{X=\{1,2,3\}}) - \Obs(\Phi_{0})\bigg]\,.\nn\\
      & + \textrm{inclusive power corrections in }\Tau_{0,\delta} \nn\,,
\end{align}
\endgroup
where the inclusive NNLO cross-section at fixed underlying-Born kinematics  in the second and last line above, $\df
\sigma_{1}^{\delta\textrm{NNLO}} / \df \Phi_{1} \ \Obs(\Phi_0)$, can be obtained by projecting $\Phi_{X=\{1,2,3\}}\to \Phi_0$ and evaluating the observable on the $\Phi_0$ configuration in \eq{finalNNLO}, giving
\begin{align}
  \label{eq:finalNNLOone}
\frac{ \df
          \sigma_{1}^{\delta\textrm{NNLO}}}{\df \Phi_{1}}\ \Obs(\Phi_0) = &  \Obs(\Phi_0)\
                                                                            \Bigg\{
                                                                            \frac {\df \Sigma_{N,\textrm{sub.}}^{\delta\textrm{NNLO}}}{\df \Phi_1} \big( \Tau_1^{\textrm{cut}}\big)   \\
	&  + \int  \frac {\df \Phi_{2}}{\df \Phi_{1}} \bigg [ { \frac{ \df \sigma_{2}^{\delta\textrm{NLO}}}{\df \Phi_{2}}}  - \frac {\df \sigma_{1,\textrm{sub.}}^{\delta\textrm{NNLO}}}{\df \Phi_1\,\df \Tau_1} \ P(\Phi_{2}) \, \theta\big( \Tau_1 (\Phi_{2}) < \Tau_1^{\textrm{cut}}\big)  \bigg]  \nn \\
	& \qquad \qquad \qquad \times  \ \theta \Big( \Tau_1(\Phi_{X=\{2,3\}}) > \Tau_{1,\delta} \Big)\nn \\
      & + \textrm{inclusive power corrections in }\Tau_{1,\delta} \Bigg \}\nn\,.
\end{align}
The dependence on the observable must instead be retained during the
evaluation of the $\df \sigma_{1}^{\delta\textrm{NNLO}} / \df \Phi_1\ \Obs(\Phi_{X=\{1,2,3\}})$ contributions appearing in the last line of \eq{finalN3LO}, in order to properly account for the fiducial power corrections.
The implementation details, the numerical dependence on the neglected power
corrections in both $\Tau_{0,\delta}$ and $\Tau_{1,\delta}$ and the results
will be discussed in a future publication.

%%%%%%%%%%%%%%%%%%%%%%%%%%%%%%%%%%%%%%%%%%%%%%%%%%%%%%%%%%%%%%%%%%%%%%%%%%%%%%%

\section{Conclusions}
\label{sec:conclusions}
%%%%%
In this work, we present a novel calculation of fully-exclusive NNLO QCD corrections, which employs genuine nonlocal subtractions based on $N$-jettiness resolution variables, in combination with the Projection-to-Born (P2B) method, to effectively handle fiducial power corrections (FPCs) in a fully-differential context.  We show numerical results for both colour-singlet and colour-singlet+jet processes within the \geneva{} framework.
To the best of our knowledge, this represents the first implementation of such a nonlocal subtraction method within a general-purpose event generator and the first application of the P2B method for processes with final-state jets which are divergent at the Born level.

We have carefully addressed the challenges associated with the usage of a
dynamic resolution cutoff, which is necessary to correctly handle the
complications associated with multi-scale problems, providing prescriptions for implementing subtraction terms that preserve phase-space constraints and accommodate dynamic cuts. We demonstrated the robustness and accuracy of our implementation through detailed studies of neutral Drell–Yan and $Z$+jet production at the LHC, validating our results against those obtained from \nnlojet{}. The inclusion of FPCs further improves the agreement in fiducial regions, underscoring the importance of these effects in precision collider phenomenology.
Our results show that the present subtraction-based implementation yields marginally better numerical convergence and lower integration errors than traditional slicing approaches, even under demanding kinematic constraints. Furthermore, the ability to incorporate the leading next-to-leading-power logarithms in the subtractions allows for more stable predictions with less sensitivity to the infrared resolution parameter.

Looking ahead, our approach lays the groundwork for future extensions. In particular, the framework presented here is well-suited for the computation of fully-differential N$^3$LO corrections for colour-singlet production at hadron colliders, which are of paramount importance for matching the precision of the upcoming HL-LHC data. Moreover, the formalism can be naturally extended to processes with higher jet multiplicities or to include electroweak corrections.
We anticipate that this work will be a key step toward the consistent inclusion of NNLO (and beyond) corrections within NNLO+PS event generators for processes with final-state jets, allowing for more precise theoretical predictions and an improved understanding of QCD dynamics at high-energy colliders.
\\~\\
\noindent {\bf Acknowledgments}

We are grateful to F.~Tackmann, J.K.L.~Michel, G.~Vita and A.~Huss for discussions. We also thank our \geneva collaborators   A.~Gavardi, M.A.~Lim, D.~Napoletano and G.~ Marinelli for valuable exchanges  and their work on the \geneva{} code.
We acknowledge financial support, supercomputing resources and support from ICSC – Centro
Nazionale di Ricerca in High Performance Computing, Big Data and Quantum Computing –
and hosting entity, funded by European Union – NextGenerationEU.

\clearpage

\addcontentsline{toc}{section}{References}

\bibliographystyle{jhep}
\bibliography{geneva}
\end{document}